\documentclass[11pt]{article}
\pdfoutput=1

\usepackage{jheppub}
\usepackage{amssymb,amsfonts,amsmath,amsthm,lineno}
\usepackage{enumerate}
\usepackage{mathrsfs}
\usepackage{bbold}
\usepackage{tikz}
\usepackage{hyperref}
\usepackage{slashed}
\usepackage{xargs}
\usepackage{tikz}

\makeatletter
\newcommand{\Vast}{\bBigg@{4.75}}
\makeatother

\definecolor{vub}{RGB}{0,52,154}
\definecolor{orange}{rgb}{1.0, 0.5, 0.0}
\usepackage[colorinlistoftodos,prependcaption,textsize=tiny]{todonotes}
\newcommandx{\unsure}[2][1=]{\todo[linecolor=vub,backgroundcolor=vub!25,bordercolor=vub,#1]{#2}}
\newcommandx{\fn}[2][1=]{\todo[linecolor=orange,backgroundcolor=orange!25,bordercolor=orange,#1]{#2}}

\newcommand{\dd}{\mathrm{d}}

\newcommand{\be}{\begin{equation}}
\newcommand{\ee}{\end{equation}}
\newcommand{\bea}{\begin{eqnarray}}
\newcommand{\eea}{\end{eqnarray}}
\newcommand{\de}{\text{d}}

\newcommand{\CF}{\mathcal{F}}

\newcommand{\CM}{\mathcal{M}}

\newcommand{\CQ}{\mathcal{Q}}

\newcommand{\CT}{\mathcal{T}}

\newcommand{\CV}{\mathcal{V}}

\newcommand{\lr}{\left (}
\newcommand{\rr}{\right )}

\newcommand\qt\tau

\newcommand{\p}{\partial}

\renewcommand{\tilde}[1]{\widetilde{#1}}
\newcommand{\tr}{\text{tr}}

\makeatletter
\renewcommand{\@seccntformat}[1]{\csname the#1\endcsname.\,\,}
\makeatother
\allowdisplaybreaks

\let \savenumberline \numberline
\def \numberline#1{\savenumberline{#1.}}

\makeatletter
\def\@fpheader{\relax}
\makeatother

\def\bea{\begin{eqnarray}}
\def\eea{\end{eqnarray}}

\usetikzlibrary{shapes,arrows,chains}
\usetikzlibrary{decorations.markings}
\usetikzlibrary{decorations.pathmorphing}
\usetikzlibrary{shapes.multipart}
\tikzset{snake it/.style={decorate, decoration=snake}}

\title{Axions, Three-Forms, and M-Theory}
\author[a]{Florian Niedermann}
\author[a, b]{and Ziqi Yan\medskip}
\emailAdd{florian.niedermann@su.se}
\emailAdd{ziqi.yan@nbi.ku.dk}
\affiliation[a]{Nordita, KTH Royal Institute of Technology and Stockholm University\\
Hannes Alfv\'{e}ns v\"{a}g 12, SE-106 91 Stockholm, Sweden \smallskip}
\affiliation[b]{Center of Gravity, The Niels Bohr Institute, Copenhagen University\\ 
Blegdamsvej 17, DK-2100 Copenhagen \O, Denmark 
}
\abstract{
Scalar fields with masses protected by global shift symmetries, commonly referred to as axions, are abundantly used in effective field theories in cosmology and particle physics. However, global symmetries cannot be expected to be protected at the fundamental level. Finding consistent ultra-violet completions for axions is therefore a necessity. 
In this work, we identify the axion with the position mode of a charged 3-brane in (4+1)-dimensions. The shift symmetry of the axion is then a residual diffeomorphism in the fifth dimension orthogonal to the brane. 
Meanwhile, the brane is coupled to a flux in the fifth dimension. From the (3+1)-dimensional perspective, this construction generates  (perturbatively) a mass for the axion and matches previously known proposals in the literature based on the coupling between the axion and a three-form gauge field. In a second step, we uplift this (4+1)-dimensional model to M-theory, where the same three-form is found to couple to the membrane with a (2+1)-dimensional worldvolume. In particular, our proposal also elucidates the duality between the axion and a two-form gauge field 
in the literature. We show that this dual two-form couples to the boundary of an open membrane in M-theory. Finally, we comment on the relations to and differences from other closed and open string axion monodromy models. 
}  

\begin{document}
\begin{flushright}
NORDITA 2025-045
\end{flushright}

\maketitle

\section{Introduction}

Light scalar fields are ubiquitous in cosmological model building. For example, ultralight scalar fields are prominent candidates for dark matter with unique signatures across a broad range of observable scales~\cite{Preskill:1982cy, Dine:1982ah, Abbott:1982af} 
(for recent phenomenological reviews see~\cite{Ferreira:2020fam, Niemeyer:2019aqm}).  
Moreover, even lighter scalar fields could explain the observed accelerated expansion of the Universe. In this case, commonly referred to as quintessence models~\cite{Ratra:1987rm, Wetterich:1987fm, Copeland:2006wr, Bhattacharya:2024kxp,Kaloper:2008qs}, the flatness of the potential is needed to make the scalar field slowly evolve on cosmological time scales. There has been a renewed interest in this type of model building~({including axion constructions, \it e.g.}~\cite{Cicoli:2021skd, Cicoli:2024yqh, Borghetto:2025jrk, Khoury:2025txd}), sparked by recent results from the Dark Energy Spectroscopic Instrument indicating that dark energy may be slowly evolving~\cite{DESI:2025zgx}. Similarly, different models that address the Hubble tension, a discrepancy in the measured value of the expansion rate of the Universe~\cite{Riess:2021jrx}, rely on an early phase of dark energy arising from the potential energy of (ultra)light scalar fields~\cite{Poulin:2018cxd,Niedermann:2019olb, Cruz:2023lmn, Chatrchyan:2024xjj}. 
In a different context, the flatness of their potential also makes light scalars attractive for inflation. Here, the (relative) smallness of the mass parameter is required by the observed near-scale invariance of the primordial scalar power spectrum~\cite{Freese:1990rb}. Models of axion monodromy provide a prominent example~\cite{Silverstein:2008sg, McAllister:2008hb} (see also~\cite{Kaloper:2008fb, Kaloper:2011jz}). The list of possible examples does of course not end here ({\it e.g.}~\cite{Dvali:2003br, Graham:2015cka, Domcke:2020kcp,Kaloper:2023kua}).

A common theoretical challenge with light scalars is that, in an effective field theory (EFT) context, their masses often receive large quantum corrections that scale quadratically with the cutoff of the theory, making them sensitive to unknown ultra-violet (UV) physics. A typical way out is to resort to the principle of technical naturalness and seek an underlying symmetry. One possibility -- employed by most of the models mentioned above -- is to identify the scalar field with the pseudo Nambu-Goldstone boson associated with a spontaneously broken symmetry that is non-linearly realized as a continuous or discrete shift symmetry. 
We will follow the usual convention and collectively refer to scalar fields of this type as axions.\,\footnote{The name `axion' is more often used as an umbrella word for such light pseudo Nambu-Goldstone bosons in more string theory-oriented literature, while in the phenomenological literature they are commonly referred to as ALPs (axion-like particles).} 
The simplest non-linear completion, first proposed by Peccei and Quinn~\cite{Peccei:1977hh} in the context of the quantum chromodynamics (QCD) axion, is to identify the axion with the phase of a complex scalar field that is charged under a \textit{global} U(1) symmetry. If this symmetry exhibits a small explicit breaking, the scalar field receives a small mass. In this case, the mass is protected against quantum corrections because there is an enhanced symmetry in the massless limit. However, this reasoning fails once quantum gravity is considered, as it is widely believed that gravitational effects generically violate global symmetries. This problem, also known as the axion `quality' problem, makes it necessary to think about alternative UV embeddings of the axions that rely on \textit{local} rather than global symmetries.

In this paper, we propose an axion scenario based on a simple geometric picture where the axion's shift symmetry is the residual part of a diffeomorphism, with potential applications to cosmological models in a broad sense. In a second part, we provide an uplift to string and M-theory. Our work is motivated by a reformulation of the axion in a dual frame put forward by Dvali~\cite{Dvali:2005an, Dvali:2022fdv}. In this dual frame, the axion is mapped in a standard way to an antisymmetric two-form gauge potential, which has a \textit{local} symmetry. A mass of the axion is then generated by coupling the dual two-form to an extra three-form gauge potential.\,\footnote{This construction, which can be understood as a Higgs mechanism giving mass to the three-form gauge potential, was already discussed in early work such as~\cite{Aurilia:1980jz}.} Importantly, in the dual frame, one only observes gauge symmetries, which are redundancies in the formulation, suitable for a possible self-consistent embedding in quantum gravity. A closely related construction for generating the axion mass was then used by Kaloper, Lawrence, and Sorbo in~\cite{Kaloper:2008fb, Kaloper:2011jz} for an EFT formulation of axion monodromy. We will show that our proposal provides an explicit geometric interpretation as a five-dimensional field theory for these previous works, as well as an embedding in string theory.

In view of the higher-form reformulation of the axion theory, it is natural to wonder: 
%\vspace{2mm}
%\leftskip1cm\relax
%\rightskip1cm\relax
%\noindent 
\emph{Does there exist a more fundamental theory of branes that underlies the dual formalism of the axion in terms of higher forms, and hence the axion itself?} 
%\leftskip0cm\relax
%\rightskip0cm\relax
%\vspace{2mm}
\begin{figure}
    \centering    \includegraphics[scale=0.22]{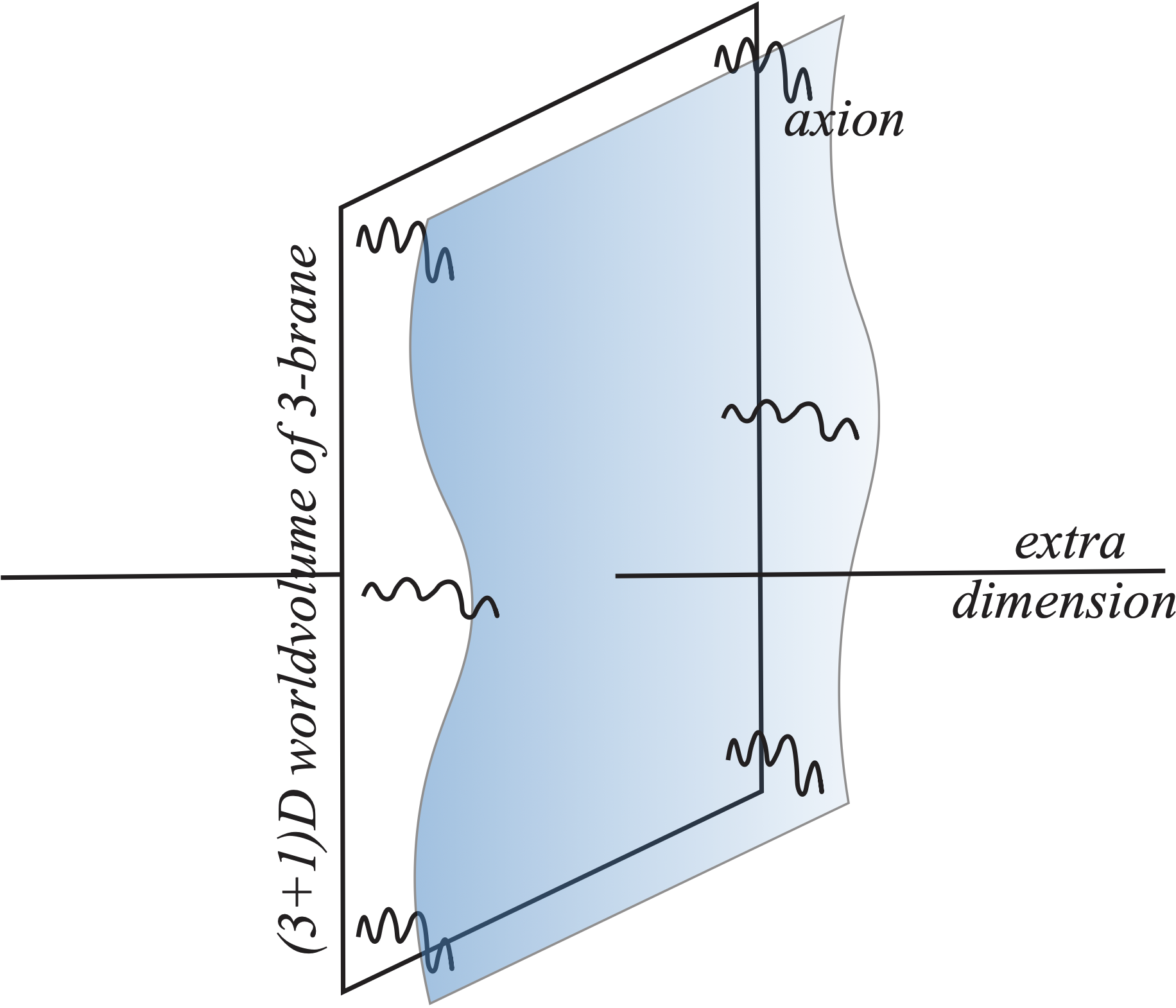} 
    \caption{Axion as a brane-bending mode from the shape of a 3-brane in the fifth extra dimension, which is compactified on a circle.}
    \label{fig:axionuniverse}
\end{figure}
%
%\noindent 
We build a physical model of a universe with a 3-brane to address this question. 
We will first describe how this works for the axion itself before delving into its dual formalism, which is more involved but, nevertheless, essential for revealing a more fundamental understanding of this scenario. 
We will start with a toy model, where we consider a 3-brane embedded in a (4+1)-dimensional spacetime, with the extra spatial direction orthogonal to the brane. We then require that this fifth dimension is compactified on a circle, and identify the position mode that perturbs the shape of the 3-brane as the axion field. See Figure~\ref{fig:axionuniverse} for a sketch. In this scenario, the global U(1) symmetry of the axion field is geometrized to be the isometry along the compactification via the Stueckelberg trick. The global U(1) is just part of the diffeomorphism symmetry that survives after fixing the position of the brane in the compact dimension. Therefore, it is not surprising that the U(1) becomes manifestly a gauge symmetry in the dual frame of the two-form.\,\footnote{This is reminiscent of T-duality in string theory, where the position mode is T-dual to a component of a gauge potential. See also~\cite{Dine:1988nrl} for further relevant discussions.} This explains how the global U(1) could be conciliated with quantum gravity. The identification of the axion with a position modulus also implies that it picks up a sign under a higher-dimensional mirror transformation, and accordingly the pseudo-scalar character of the axion is inherited from the parity invariance of the higher-dimensional theory.

The novelty of our construction of the axion as a brane position mode lies in that the axion mass arises from the coupling of the 3-brane to a four-form gauge potential when the  corresponding flux field winds around the compact dimension.\,\footnote{We emphasize that there are other models where the axion/inflaton is identified with the position mode/Wilson line of a brane in string theory, e.g.~\cite{Dasgupta:2002ew, Avgoustidis:2006zp, Kachru:2003sx, Hebecker:2014eua, Marchesano:2014mla}. A more phenomenological example is the extranatural inflation~\cite{Arkani-Hamed:2003xts}, which is a five-dimensional model that provides a top-down perspective on the pseudonatural inflation~\cite{Arkani-Hamed:2003wrq}, where the axion mass is generated by one-loop quantum corrections.}  Upon dimensional reduction, this explains the presence of the three-form gauge potential in four dimensions. It also  admits a simple expression of the axion mass in terms of the size of the compact dimension, brane tension, and charge [see Eq.~\eqref{eq:axion_mass}].

\definecolor{lb}{rgb}{0.07, 0.6, 1} 

\begin{figure}
    \centering
    \hspace{8mm}
    \includegraphics[scale=0.8]{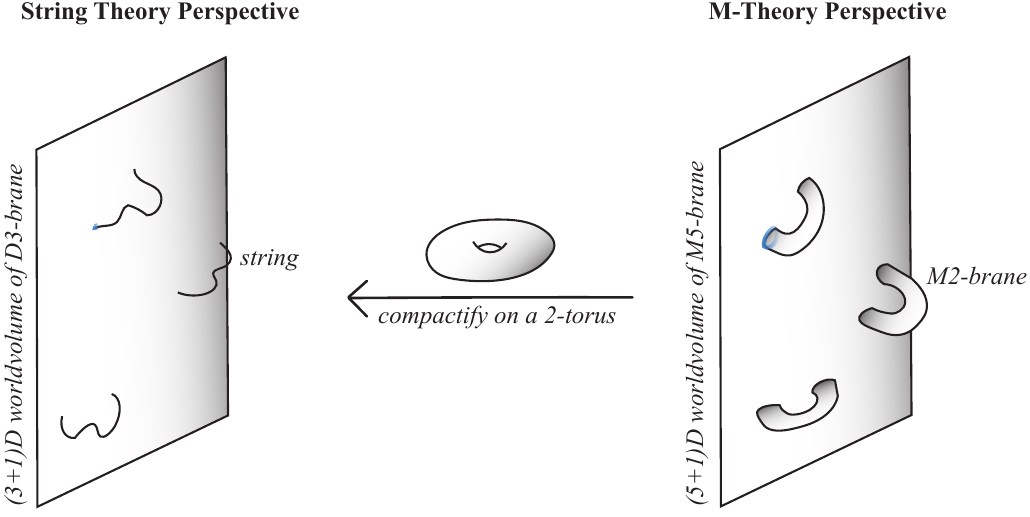} \\[-123pt]
    \hspace{-8mm}
    \begin{tikzpicture}
        \node at (-9.3,0.5) {\scalebox{0.8}{\color{lb}\emph{axion}}};
        \node at (0.8,0.8) {\hspace{-2.2cm}\scalebox{0.8}{\color{lb}$\mathbb{A}^{\!\text{\scalebox{0.8}{$(2)$}}}_{\phantom{\dagger}}$}};
    \end{tikzpicture}\\[100pt]
    \caption{The string and M-theory uplift of the axion universe. From the string theory perspective, the axion is an open string mode that perturbs the shape of the D3-brane. From the M-theory perspective, the axion is dualized to be a differential two-form potential $\mathbb{A}^{\!(2)}$, and it is coupled to open M2-branes that end on an M5-brane.}
    \label{fig:smt}
\end{figure}
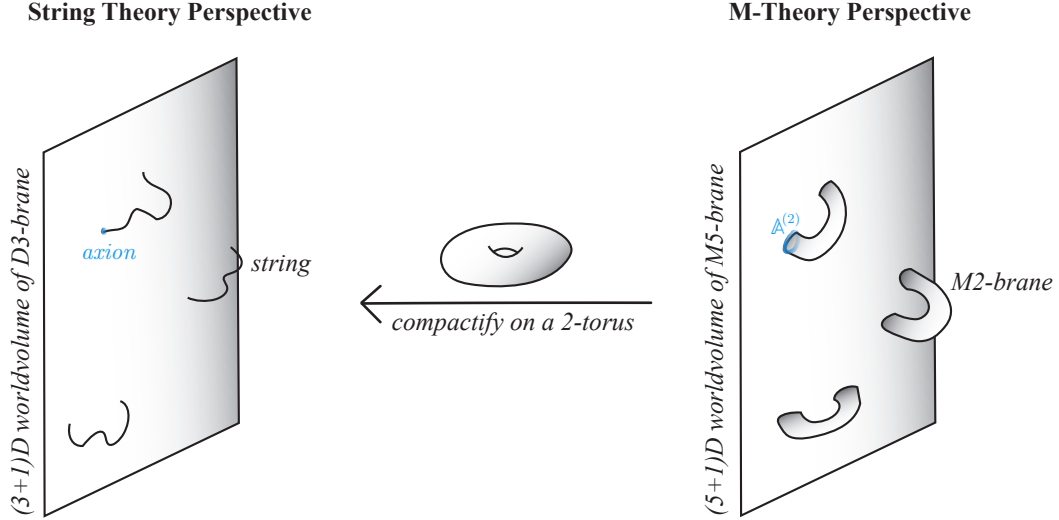

We will then take this idea one step further and consider how our proposal can be embedded within the framework of string and M-theory, where branes naturally appear.\,\footnote{This embedding -- regardless of its phenomenological relevance -- provides further insights for the construction. In fact, the authors of this paper were originally led to this brane construction by following the inspiration from string and M-theory.} 
Axions are considered as a generic prediction of string theory, giving rise to what is sometimes referred to as the `string axiverse'~\cite{Arvanitaki:2009fg}. However, in the literature, the dual two-form is usually interpreted as the Kalb-Ramond (or Ramond-Ramond) field coupled to the closed string, just like how the electromagnetic one-form potential is coupled to a particle. This is the idea behind \emph{e.g.}~standard axion monodromy, where the axion mass is then generated via an instantonic string wormhole~\cite{Kallosh:1995hi}. Conversely, in our proposal, \emph{no} such Kalb-Ramond form is needed. Instead, the dual two-form finds a natural interpretation when we uplift the construction all the way to M-theory in eleven dimensions: it is the two-form gauge potential coupled to the open membrane! It is then a logical consequence that the extra three-form gauge potential, responsible for the mass generation, is coupled to the closed membrane. This means that our toy-model universe has to be uplifted to an M-theory five-brane with a (5+1)-dimensional worldvolume. After compactifying the five-brane on a two-torus, we are led to the string theory perspective, where we recover the five-dimensional toy model that we have explained earlier, with our universe being a 3-brane and the axion a brane-position mode. These two perspectives from string theory (for the scalar axion) and M-theory (for the dual two-form) are precisely related to each other via a non-perturbative duality transformation~\cite{Berman:1998va, Berman:1998sz}. This embedding in string and M-theory points towards one possible non-linear completion in terms of gauge potentials that recovers the effective Goldstone theory of the axion after spontaneous symmetry breaking, where the relations between different descriptions are manifested as string dualities. This string theoretical perspective also shows that the axion we propose here is precisely \emph{not} the QCD axion. See Figure~\ref{fig:smt} for an illustration. 

Our proposal of the axion as a brane-position mode is in spirit reminiscent of the open string axion monodromy considered in~\cite{Marchesano:2014mla}, where the axion arises from a massive Wilson line. As Wilson lines are T-dual to brane locations in string theory, these are two complementary perspectives of open string axions. However, the mass generation mechanism used in~\cite{Marchesano:2014mla} involves a twisted torus and differs from our work. 

This paper is organized as below. We start in Section~\ref{sec:eftp} with formulating our five-dimensional toy model from an EFT perspective, where \emph{no} background knowledge of string theory is required. In Section~\ref{sec:mtu} we explain how the EFT can be uplifted to M-theory, where we give a pedagogical introduction to all the essential ingredients from string and M-theory that are required to understand the essence of the construction, with the hope that the materials there are also accessible to readers with a more phenomenological background. Both the EFT and M-theoretical parts of the paper can be read independently, depending on the interest of the reader. In Section~\ref{sec:concl} we conclude the paper and outlook some future directions. In Appendix~\ref{sec:ccsam} we discuss differences and relations between our model and previous constructions of both closed and open string axions. 

%%%%%%%%%%%%%%%%%%%%%%%%%%%%%%%
\section{Effective Field Theory Perspective} \label{sec:eftp}
%%%%%%%%%%%%%%%%%%%%%%%%%%%%%%%

In this section we present our EFT construction for the proposed axion as a brane-position mode. We highlight that the U(1) global symmetry of the axion is a remnant of the diffeomorphism that is a part of the purported quantum gravity at very high energies. We then dualize the axion on the four-dimensional manifold of the 3-brane to the frame where the axion is replaced with a two-form potential. Finally, we discuss the mass generation from an extra three-form potential and how it arises from the higher-dimensional theory. Through these discussions, we will review the formulations in~\cite{Dvali:2005an,Kaloper:2008fb, Kaloper:2011jz} while adding in our new perspectives from the brane-world model. 

\subsection{Axions from Brane Bending} \label{sec:afbb}

Consider our Universe as a three-dimensional brane embedded within a five-dimensional ambient spacetime, and compactify the extra fifth dimension that is orthogonal to the brane over a circle with radius $R$\,. Denote the brane worldvolume as $\CM_4$ and denote the five-dimensional ambient spacetime as $\CM_5$\,. We introduce the brane worldvolume coordinates $x^\mu$ with $\mu = 0\,, \, \cdots, \, 3$ and the spacetime coordinates $X^\text{A}$ with $\text{M} = 0\,, \, \cdots, \, 4$\,. 
Define $f^\text{M} (x)$ to be the embedding function that describes how the brane is embedded within $\CM_5$\,, such that
\be \label{eq:embedding}
    X^\text{M} \Big|_{\CM_4} = f^\text{M} (x)\,,
\ee
For simplicity, we focus on flat worldvolume and flat ambient spacetime. We require that the brane dynamics be described by the pure-tension action,\,\footnote{We will see later in Section~\ref{sec:tcmfb} that this action is part of the D3-brane worldvolume theory in type IIB superstring theory.}
\be \label{eq:bywwa}
    S = - \mathcal{T} \int_{\CM_4} \de^4 x \, \sqrt{- \det \Bigl( \p^{}_\mu f^\text{M} \, \p^{}_\nu f^\text{N} \, \eta^{}_\text{MN} \Bigr)}\,.
\ee
Here, $\CT$ is the brane tension. 
We focus on the brane-position mode that perturbs the shape of the brane along the fifth spacetime direction. In the static gauge, we align the brane within the $X^\mu$ directions and assume that the brane is located at $y^{}_0 = \mathrm{const}$ in $X^4$\,. The embedding function $f^\text{M}$ is then evaluated to be
\be \label{eq:embf}
    f^\mu = x^\mu\,,
        \qquad%
    f^4 = y^{}_0 + R \, \pi (x)\,. 
\ee
The scalar field $\pi(x)$ is the brane-position mode that dynamically changes the shape of the brane. This excitation can be thought of as a Nambu-Goldstone boson from spontaneously breaking the translational symmetry in the compactified fifth direction by localizing the brane there.   
Moreover, the action \eqref{eq:bywwa} is invariant under five-dimensional parity transformations, $(X^i\,,\, X^4) \to (-X^i\,,\, -X^4)$\,, where $X^i$ with $i = 1\,, \, 2\,, \, 3$ are the three spatial directions along which the brane extends and $X^4$ is the extra fifth direction. 
Due to \eqref{eq:embedding} and \eqref{eq:embf}, this corresponds in the worldvolume theory of the brane to the pseudo-scalar transformation $\pi(t\,,\,x^i) \to -\pi(t\,,\,-x^i)$\,, where $t$ is the worldvolume time and $x^\mu = (t\,, \, x^i)$\,. In other words, this five-dimensional parity  manifests itself in the 4D theory through the pseudo-scalar character of the brane-position field $\pi(x)$, making it a natural axion candidate.
In fact, we will see that this construction provides a clear physical realization of the axion theory proposed in~\cite{Dvali:2005an,Dvali:2022fdv}, which supposedly solves the quality problem.

Considering small brane displacements, we only keep the leading term from the brane action~\eqref{eq:bywwa} that is quadratic in $\pi (x)$\,, which takes the simple form
\be \label{eq:fa}
    S^{(2)}[a] = - \frac{1}{2} \int \de^4 x \, \p_\mu a(x) \, \p^\mu a(x) \,,
\ee
where we introduced the canonically normalized field 
\be \label{eq:ax}
    a(x) = R \, \sqrt{\mathcal{T}} \, \pi (x)\,.
\ee
This is the four-dimensional QFT of a single free scalar boson, which is invariant under a global shift, \emph{i.e.} $a \rightarrow a + \text{const}$. However, we have to keep in mind that this global shift came from fixing the location of the brane to be at $y^{}_0$ in the fifth direction. We can choose to describe the embedding of the brane within $\CM_5$ in a more covariant way such that the location of the brane is left unfixed. This is implemented by introducing a $U(1)$ gauge potential $b_\mu$ and rewriting the free action~\eqref{eq:fa} equivalently as
\be \label{eq:gauo}
    S^{(2)}_\text{gauged} \bigl[a\,, \, b^{(1)}, \, \mathbb{A}^{(2)} \bigr]  = - \frac{1}{2} \int \dd^4 x \, \Bigl( D_\mu a \, D^\mu a + 
    \epsilon^{\mu\nu\rho\sigma} \, \mathbb{A}_{\mu\nu} \, \p_\rho b_\sigma \Bigr)\,.
\ee
Here, the covariant derivative $D_\mu$ acts as $D_\mu a = \p_\mu a + b_\mu$ and the Levi-Civita symbol $\epsilon^{\mu\nu\rho\sigma}$ satisfies $\epsilon^{0123} = 1$\,. The Lagrange multiplier\footnote{Later in the text, we will also use analogous notation with a superscript to denote other higher-form fields.} $A^{(2)} = \frac{1}{2} \, \mathbb{A}_{\mu\nu} \, \dd x^\mu \wedge \dd x^\nu$ is an antisymmetric two-form field, which imposes the condition $\p_{[\rho} b_{\sigma]} = 0$\,. Later, we will see that this two-form replaces the axion in the dual language. Locally, this constraint implies that $b_\mu$ is exact, \emph{i.e.}, 
\be \label{eq:amdy}
    b_\mu (x) = \sqrt{\mathcal{T}} \, \p_\mu y (x)\,.
\ee
Plugging Eq.~\eqref{eq:amdy} back into the action~\eqref{eq:gauo} yields
\be \label{eq:fa_gen}
    S^{(2)}[a] = - \frac{\mathcal{T}}{2} \int \de^4 x \, \p_\mu \!\! \lr y + \frac{a}{\sqrt{\CT}} \rr \, \p^\mu \!\! \lr y + \frac{a}{\sqrt{\CT}} \rr, 
\ee
which could have been obtained directly from~\eqref{eq:bywwa} by using an embedding with arbitrary brane position, \emph{i.e.}\ $f^4 = y(x) + a (x) / \sqrt{\CT}$\,. 
The action~\eqref{eq:gauo} is invariant under the following $U(1)$ gauge transformation,
\be\label{eq:sym_U1}
    a (x) \rightarrow a (x) + \sqrt{\CT} \, \xi(x)\,,
        \qquad%
    b_\mu (x) \rightarrow b_\mu (x) - \sqrt{\CT} \, \p_\mu \xi (x)\,. 
\ee
which `gauges' the isometry and recovers the diffeomorphism $X^4 \to X^4 + \xi(x)$ along the extra dimension~$X^4$ due to Eq.~\eqref{eq:amdy}.  
Note that the Lagrange multiplier term in Eq.~\eqref{eq:gauo} is indispensable as it ensures that $b_\mu$ is pure gauge. Therefore, the gauge potential $b_\mu$ does \emph{not} introduce any extra physical degrees of freedom. 

We now fix the gauge parameter $\xi$ in Eq.~\eqref{eq:sym_U1} such that $\p_\mu \xi = \p_\mu y$\,, in which case $b_\mu$ is set to zero due to Eq.~\eqref{eq:amdy} and we recover the original action~\eqref{eq:fa}. Crucially, the gauge-fixing condition $\p_\mu \xi = \p_\mu y$ leaves a global shift as a \textit{residual} gauge symmetry: this gauge-fixing condition is preserved under $y \rightarrow y + \text{c}$\,, when c is constant. It is this residual symmetry that we identify with the usual axion shift symmetry. However, at a more fundamental level, it is inherited from the diffeomorphism in the fifth dimension. Geometrically, it corresponds to a translation of the brane in the fifth dimension. 

At face value, this discussion shows that the global shift symmetry of the brane-position mode is robust under quantum gravity corrections. To make a connection to the axion (or axion-like particles), we also need to generate a mass term. In order to facilitate this mass generation, we will first show that Eq.~\eqref{eq:gauo} naturally has
a dual formulation in terms of a two-form that will be identified with $\mathbb{A}^{(2)}$\,. We will discuss this dualization in the next subsection, which will play a central role later in the embedding of our proposal in M-theory. 

Note that our proposal is distinct from a gauged version of the Peccei-Quinn theory. We reiterate that the gauging procedure leading to the action~\eqref{eq:gauo} only introduces a gauge potential $b_\mu$ that does not carry any physical degrees of freedom. This is distinct from the standard gauging in the context of the Abelian Higgs mechanism, where the gauge boson carries three degrees of freedom after having `eaten' the scalar mode. Instead, the gauge symmetry in the action~\eqref{eq:gauo} is purely of the Stueckelberg type. We also emphasize that, in this construction, $a(x)$ is not the phase of a complex scalar field as in the case of the Peccei-Quinn theory. 

%%%%%%%%%%%%%%%%%%%%%%%%%%%%%%%%%%%%%%%%%
\subsection{Two-Form Potential from Dualizing the Axion} \label{sec:catfd}
%%%%%%%%%%%%%%%%%%%%%%%%%%%%%%%%%%%%%%%%%

There is another way of making the gauge origin of the axion's constant shift symmetry manifest, more commonly highlighted in the literature~\cite{Quevedo:1996uu, Dvali:2005an,Kaloper:2016fbr,Dvali:2022fdv, Burgess:2023ifd, Anber:2024gis}. In short, the idea is to use the fact that, in four dimensions, a scalar field $a(x)$ is dual to a two-form gauge potential. Here, we will argue that this statement is complementary to the geometrical picture introduced above in Section~\ref{sec:afbb}. In particular, we will show that this dual two-form can be identified with~$\mathbb{A}^{(2)}$ in the gauged action~\eqref{eq:gauo}.

We start with describing the dual frame in terms of the two-form gauge potential $\mathbb{A}^{(2)}$\,, and then make connection to the scalar theory~\eqref{eq:fa}.  
Consider the free action
\be \label{eq:S_2form}
    S^{(2)}_\text{dual}\bigl[\mathbb{A}^{(2)}\bigr] = - \frac{1}{2 \cdot 3!} \int_{\CM_4} \dd^4 x \, \mathbb{F}^{\mu\nu\rho} \, \mathbb{F}^{}_{\mu\nu\rho}\,,
\ee
where we introduced the field strength $\mathbb{F}^{(3)}=\mathrm{d}\mathbb{A}^{(2)}$\,, or, more explicitly, 
\be
    \mathbb{F}^{}_{\mu\nu\rho} = \p^{}_\mu \mathbb{A}^{}_{\nu\rho} + \p^{}_\nu \mathbb{A}^{}_{\rho\mu} + \p^{}_\rho \mathbb{A}^{}_{\mu\nu}\,.
\ee
This action is invariant under the gauge transformation 
\begin{align}\label{eq:sym_2form}
\mathbb{A}^{(2)} \to \mathbb{A}^{(2)} + \mathrm{d}\lambda^{(1)}
\end{align}
for a general one-form $\lambda^{(1)} = \lambda_\mu \, \mathrm{d}x^\mu$. 
We now introduce a new action that is equivalent to Eq.~\eqref{eq:S_2form}, but with $\mathbb{F}^{(3)}$ (rather than $\mathbb{A}^{(2)}$) being the fundamental field. This is achieved by introducing a Lagrange multiplier field $a_\mu$ and then defining a `parent' action,
\be \label{eq:sparent_EFT}
    S_\text{parent} \bigl[\mathbb{A}^{(2)},\mathbb{F}^{(3)}\,, \, a^{(1)} \bigr]  = - \frac{1}{2 \cdot 3!} \int_{\CM_4} \dd^4 x \, \mathbb{F}^{\mu\nu\rho} \, \mathbb{F}^{}_{\mu\nu\rho} 
    + S_\text{g.f.}\,,
\ee
with the generating functional,
\be \label{eq:gfworr_EFT}
    S_\text{g.f.} = \frac{1}{3!} \, \int_{\CM_4} \dd^4 x  \, \epsilon^{\mu\nu\rho\sigma} \, a^{}_\mu \, \Bigl( \mathbb{F} - \dd\mathbb{A} \Bigr){}^{}_{\nu\rho\sigma}\,. 
\ee
Integrating out $a_\mu$ in the path integral associated with the parent action~\eqref{eq:sparent_EFT} imposes the condition $\mathbb{F}^{}_{\mu\nu\rho} = \bigl( \dd\mathbb{A} \bigr){}^{}_{\mu\nu\rho}$\,, which leads us back to the original action~\eqref{eq:S_2form}. Instead, we now integrate out $\mathbb{A}_{\mu\nu}$\,, 
which imposes the exactness condition $\dd a^{(1)} = 0$\,, where $a^{(1)} = a_\mu \, \dd x^\mu$\,. This condition can be solved locally
by
\be \label{eq:pmmp}
    a_\mu = \p_\mu a\,,
\ee
where $a$ is a scalar field (later playing the role of the axion). Plugging Eq.~\eqref{eq:pmmp} back into the parent action, we obtain
\begin{align}\label{eq:S_3form}
    S^{(2)}_\text{dual}[\mathbb{F}^{(3)},a] = - \frac{1}{  3!} \int_{\CM_4} \dd^4 x \left( \tfrac{1}{2} \, \mathbb{F}^{\mu\nu\rho} \, \mathbb{F}^{}_{\mu\nu\rho} + a \, \epsilon^{\mu\nu\rho \lambda} \, \partial_\mu  \mathbb{F}_{\nu\rho\lambda} \right)\,.
\end{align}
Under the shift $a \to a + \mathrm{const}$, the action only changes by a boundary term,   
rendering it invariant in the absence of boundaries. Next, 
we note that the $\mathbb{F}^{(3)}$ integration in the path integral is a Gaussian with a saddle point at
\begin{align}\label{eq:dual1}
    \mathbb{F}_{\mu\nu\rho} = - \epsilon_{\mu\nu\rho\lambda} \, \partial^\lambda a\,.
\end{align}
Substituting this solution back into Eq.~\eqref{eq:S_3form} indeed returns the scalar field action~\eqref{eq:fa}. In the literature, this duality is typically referred to when arguing for the existence of a UV stable axion. However, we can still ask how this protection can be made more manifest in the scalar field language.  As argued before, the answer is provided by the Stueckelberg action~\eqref{eq:gauo}, which promotes the global shift symmetry to a local U(1). Notably, this Stueckelberg action already exhibits both the two-form symmetry~\eqref{eq:sym_2form} \textit{and} the U(1) symmetry~\eqref{eq:sym_U1}.

So far, we have shown how to get back the scalar field action~\eqref{eq:fa} from the dual gauge field action~\eqref{eq:S_2form} of the two-form. To make this point more explicit, we now show the reverse, \emph{i.e.}~how to get the gauge field action~ from the scalar field action, where the latter has been shown to be equivalent to the (geometrically motivated) Stueckelberg action~\eqref{eq:gauo}. To that end, we integrate out $b_\mu$ in Eq.~\eqref{eq:gauo} by using its equation of motion 
\begin{align}\label{eq:dual2}
    b_\mu = -\partial_\mu a - \frac{1}{3!} \, \epsilon_{\mu\nu\rho\sigma} \, \mathbb{F}^{\nu\rho\sigma} \,.
\end{align}
Substituting this back into \eqref{eq:gauo} indeed recovers \eqref{eq:S_2form}. We recall that integrating out $b_\mu$ does not change the number of degrees of freedom as $b_\mu$ is a pure gauge field.  If we gauge fix $b_\mu=0$ as before through a suitable choice of the diffeomorphism $\xi$ in the fifth dimension, Eqs.~\eqref{eq:dual1} and~\eqref{eq:dual2} become equivalent. In other words, Eq.~\eqref{eq:dual1} is the gauge-fixed duality map between $a$ and $\mathbb{A}^{(2)}$ where, as explained before, the axion shift symmetry can be identified as the residual global part of a local U(1).

%%%%%%%%%%%%%%%%%%%%%%%%%%%%%%%%%%%%%%%%
\subsection{Mass Generation from Three-Form Potential} \label{sec:mgtfp}
%%%%%%%%%%%%%%%%%%%%%%%%%%%%%%%%%%%%%%%%

In this section, we study a five-dimensional toy model for mass generation. A more elaborated scenario will be discussed in Section~\ref{sec:back-reaction}. 

So far, our axion field is massless. For the QCD axion, its mass arises below the QCD scale from the anomalous coupling to $F \wedge F$. In this work, we will consider a more general mechanism. It extends the approach put forward by Dvali in~\cite{Dvali:2005an} (and recently reviewed in~\cite{Kaloper:2016fbr, Burgess:2023ifd, Anber:2024gis}) where a nontrivial axion potential is generated through the coupling to a three-form field $C^{(3)}$. 
In the four-dimensional theory, the mass generating mechanism might appear rather \emph{ad hoc}. Nevertheless, in our five-dimensional model, we will see that $C^{(3)}$ arises from 
the presence of a four-form gauge field $C^{(4)}$ that naturally couples to the worldvolume of the 3-brane. We denote the associated field strength as $H^{(5)} = \mathrm{d} C^{(4)}$\,. In other words, our 3-brane now carries both a charge $\mathcal{Q}$ and tension $\mathcal{T}$. This coupling of a four-form gauge potential to a 3-brane is a higher-dimensional analog of how the electromagnetic gauge field, a one-form, couples to the worldline of a charged particle, a 0-brane.  As we will see, the four-dimensional picture of a massive scalar field then simply arises from dimensional reduction.

The five-dimensional action that generalizes \eqref{eq:bywwa} to include the coupling to the four-form gauge potential $C^{(4)}$ is
\begin{align} \label{eq:5D_full}
    S = - \frac{1}{2\cdot 5!}  \int_{\CM_5} \de^5 X \, H^{}_\mathrm{ABCDE} \, H^\mathrm{ABCDE}
    - \mathcal{T} \int_{\CM_4} \de^4 x \, \sqrt{- \det \gamma } - \mathcal{Q} \int_{\CM_4} C^{(4)}\,,
\end{align}
where the first term is the kinetic term for the four-form flux in the bulk, and the second and third term describe the brane tension and charge, respectively. We also introduced the induced metric $\gamma$, which is the pull-back of the bulk metric onto the brane,
\be
    \gamma_{\mu\nu} = \p^{}_\mu f^\text{A} \, \p^{}_\nu f^\text{B} \eta^{}_\text{AB}\,,
\ee
recovering the tension term in \eqref{eq:bywwa}. With this definition, also the charge term can be written more explicitly as the pull-back of~$C^{(4)}$\,:
\be \label{eq:qinc}
    \int_{\CM_4} C^{(4)} = \frac{1}{4!} \int_{\CM_4} \de^4 x \, 
    \epsilon^{\mu\nu\rho\sigma}\, \p^{}_\mu f^\text{A} \, \p^{}_\nu f^\text{B} \,\p^{}_\rho f^\text{C} \, \p^{}_\sigma f^\text{D} \, C_\mathrm{ABCD}(f) \,.
\ee
By substituting the brane embedding~\eqref{eq:embf} into the Wess-Zumino term~\eqref{eq:qinc}, we arrive at
\begin{align}
    \int_{\CM_4} C^{(4)} & = \frac{1}{4!} \int_{\CM_4} \de^4 x \, \epsilon^{\mu\nu\rho\sigma} \Bigl( C_\mathrm{\mu\nu\rho\sigma} + 4 \,  R \,\p^{}_\mu \pi \,  C_\mathrm{\nu\rho\sigma} \Bigr)\,, 
\end{align}
where we introduced the three-form $C_\mathrm{ \beta\gamma \delta} \equiv C_\mathrm{ 4 \beta\gamma \delta }$\,. We also used $\p_4 C_{\mu\nu\rho\sigma} = 0$\,, following from the assumption that $X^4$ is an isometry direction. We thus observe that a charged 3-brane gives rise to a linear coupling between a three-form and the brane-position mode $\pi$. 
For now, we also ignore the back-reaction of the brane on the bulk fluxes, which we will return to in the next subsection. Note that this back-reaction plays an important role in the case where the fifth dimension is compactified on a circle. However, it is also possible to instead consider a Randall-Sundrum type scenario with the fifth dimension being uncompactified.

Focusing on the terms up to quadratic order in $\pi$\,, the dimensional reduction of the five-dimensional action~\eqref{eq:5D_full} in the fifth dimension gives
\begin{align}\label{eq:dim_reduced_action}
\begin{split}
    S^{(2)} &= - \frac{1}{2} \int  \de^4 x \left( \frac{2\pi R}{4!} \, H_\mathrm{\mu\nu\rho\sigma} \, H^\mathrm{\mu\nu\rho\sigma} + \mathcal{T} R^2 \, \p_\mu \pi \, \p^\mu \pi \right) \\[4pt]
    & \quad - \frac{\mathcal{Q}}{4!} \int_{\CM_4} \de^4 x \, \epsilon^{\mu\nu\rho\sigma} \Bigl( C_\mathrm{\mu\nu\rho\sigma} + 4 \,  R \,\p^{}_\mu \pi \,  C_\mathrm{\nu\rho\sigma} \Bigr)\,,
\end{split}
\end{align}
where we introduced the field strength $H^{(4)} = \mathrm{d} C^{(3)}$. Since $C^{}_{\mu\nu\rho\sigma}$ is decoupled from the other modes in this action, we set $C^{}_{\mu\nu\rho\sigma} = 0$ below for simplicity. Rescaling the field $C^{(3)}$ such that its kinetic term is canonically normalized and using \eqref{eq:ax}, we find
\be\label{eq:S_massive}
    S^{(2)} \bigl[a\,, \, C^{(3)} \bigr]  = -\frac{1}{2} \int  \de^4 x \, \Bigl( \tfrac{1}{4!} \, H_\mathrm{\mu\nu\rho\sigma} \, H^\mathrm{\mu\nu\rho\sigma} + \, \p_\mu a \, \p^\mu a + \tfrac{2}{3
    !} \, m \, \epsilon^{\mu\nu\rho\sigma}\, \p^{}_\mu a \,  C_\mathrm{ \nu\rho\sigma} \Bigr),
\ee
where we introduced the mass scale
\be\label{eq:axion_mass}
    m = \frac{\CQ}{\sqrt{2\pi R \, \CT}}\,.
\ee
This action is invariant under the global shift  $a \rightarrow a + \text{c}$\,. 
In our construction, this is again a 
consequence of the translational symmetry of the 3-brane vacuum along the extra dimension. Moreover, it is exactly the action \eqref{eq:S_massive} that has been used previously as the starting point for giving a mass to the axion through a three-form Higgs mechanism~\cite{Dvali:2005an}. To see this explicitly, we again introduce the following parent action:
\be\label{eq:S_parent}
    S^{}_\text{parent} = -\frac{1}{2} \int  \de^4 x \, \Bigl( \tfrac{1}{4!} \, H_\mathrm{\mu\nu\rho\sigma}\, H^\mathrm{\mu\nu\rho\sigma} + \, \p_\mu a \, \p^\mu a  - \tfrac{2}{4!} \, \epsilon^{\mu\nu\rho\sigma} \,  H_\mathrm{ \mu\nu\rho\sigma} \, m \, a \Bigr) +  S_\text{g.f.}\,,
\ee
which imposes $H^{(4)} = \de C^{(3)} $ through a Lagrange multiplier $\rho$ in the generating functional
\be \label{eq:gfworr_EFT_2}
    S_\text{g.f.} = - \frac{1}{4!} \, \int_{\CM_4} \dd^4 x \, \rho \, \epsilon^{\mu\nu\rho\sigma} \Bigl( H^{(4)} - \de {C}^{(3)} \Bigr){}^{\phantom{\dagger}}_{\mu\nu\rho\sigma}\,. 
\ee
Note that $H^{(4)}$ is now treated as an independent field, and only becomes exact after integrating out $\rho$. 
Instead, we integrate out $C_{\mu\nu\rho}$ in the path integral, which imposes $\partial_\mu \rho = 0$\,. Locally, this condition is solved by setting $\rho = m \, a^{}_0$\,, with $a_0$ a constant. Plugging this solution back into the parent action~\eqref{eq:S_parent}, we obtain
\be
S^{(2)}\bigl[a\,, \, H^{(4)}\bigr]  = -\frac{1}{2} \int_{\CM_4}  \de^4 x \, \Bigl[ \tfrac{1}{4!} \, H_\mathrm{\mu\nu\rho\sigma}  H^\mathrm{\mu\nu\rho\sigma} +  \p_\mu a \, \p^\mu a  - \tfrac{2}{4!} \, \epsilon^{\mu\nu\rho\sigma} \,  H_\mathrm{ \mu\nu\rho\sigma} \, m \, \bigl(a - a_0\bigr) \Bigr].
\ee
This recovers the axion-four-form theory in \cite{Kaloper:2008fb, Kaloper:2011jz}.
Importantly, the shift symmetry of the action~\eqref{eq:S_massive} has survived as a symmetry under a \textit{simultaneous} shift of $a$ and $a_0$.\footnote{As emphasized in~\cite{Dvali:2005an,Dvali:2022fdv}, the integration constant $a_0$ carries information of the vacuum structure of \eqref{eq:S_massive}. A similar point was made in \cite{Kaloper:2016fbr}, where $a_0$ was interpreted as a `macroscopic, extensive property' of the system, drawing on an analogy with the low-energy description of a defect condensate in terms of a massive gauge field~\cite{Julia:1979ur}.}
The equation of motion from varying the parent action with respect to the independent field $H_{\mu\nu\rho}$ gives
\begin{align}\label{eq:legendre1}
    m \, \bigl( a - a_0 \bigr) = - \tfrac{1}{4!} \, \epsilon^{\mu\nu\rho\sigma} \, H_\mathrm{\mu\nu\rho\sigma}^{}\,,
\end{align}
which features a stationary point for the axion $a$ at $a^{}_0$\,. 
We then use Eq.~\eqref{eq:legendre1} to integrate out $H^{(4)}$, which gives
\be \label{eq:ma}
    S^{(2)}_\mathrm{massive}[a] = - \frac{1}{2} \int_{\CM_4} \de^4 x \, \Bigl[ \p_\mu a \, \p^\mu a + m^2 \, \bigl( a - a_0 \bigr)^2  \Bigr]\,.
\ee
The dependence on $a^{}_0$, which is related to the vacuum expectation value of the four-form field strength $H_{\mu\nu\rho\sigma}$\,, can be redefined away via the shift $a \to a + a_0$\,. This indeed leads to the action of a massive scalar field as desired. Moreover, we will see in the next section that, in a more complete setting, $a_0$ is quantized, leading to a multi-valued potential reminiscent of axion monodromy inflation. 

The explicit expression in Eq.~\eqref{eq:axion_mass} for the mass is one of the central results of this work. It shows that in our construction a small sub-Planckian axion mass can be achieved in the limit where $\mathcal{Q} \ll \sqrt{R\,\mathcal{T}} M_\mathrm{pl}$. More crucially, there is no \textit{a priori} reason to believe that such a hierarchy could not be protected by quantum gravity.

Not surprisingly, the mass generation can also be realized in the dual picture where $a$ is replaced by the two-form $\mathbb{A}^{(2)}$. To see this, we take Eq.~\eqref{eq:S_massive} as the starting point and perform the same steps taken in the massless case in Section~\ref{sec:afbb}. To be specific, we introduce  a U(1) gauge potential $b_\mu$ that makes the higher dimensional diffeomorphism invariance in the extra fifth dimension manifest by replacing $\partial_\mu a \to D_\mu a = \partial_\mu a + b_\mu$\,. With this replacement, the resulting action is indeed invariant under the Stueckelberg symmetry~\eqref{eq:sym_U1}. In addition, a two-form Lagrange multiplier field $\mathbb{A}^{(2)}$ is needed to impose the exactness condition $\p_{[\rho} b_{\sigma]} = 0$, which renders $b_\mu$ a pure gauge field as in Eq.~\eqref{eq:amdy}. Integrating out $b_\mu$ via its equation of motion \eqref{eq:dual2} then yields the dual of Eq.~\eqref{eq:S_massive}, 
\be\label{eq:fdquad0}
S_\text{dual}\bigl[\mathbb{A}^{(2)}, C^{(3)} \bigr]  = -\frac{1}{2} \! \int  \! \de^4 x \, \Bigl[ \tfrac{1}{3!} \left(\mathbb{F}^{\mu\nu\rho}- \tfrac{m}{2} \, C^\mathrm{ \mu\nu \rho} \right) \left(\mathbb{F}^{}_{\mu\nu\rho} - \tfrac{m}{2} \, C_\mathrm{ \mu\nu \rho}\right) + \tfrac{1}{4!}  H_\mathrm{\mu\nu\rho\sigma} H^\mathrm{\mu\nu\rho\sigma} \Bigr].
\ee
Again, this recovers known expressions in the literature~\cite{Burgess:2023ifd, Dvali:2022fdv,Kaloper:2016fbr}.

Note that, in the proposal in~\cite{Dvali:2005an}, the EFT was constructed specifically for the QCD axion.\footnote{For a different type of model that relaxes $\theta$ in discrete steps but still relies on a four-form formulation, see~\cite{Kaloper:2025wgn,Kaloper:2025upu}.} Even though QCD preserves C(harge conjugation), P(arity), and T(ime reversal) transformation at perturbative orders, the quantum effects of the instanton solution of QCD lead us to introduce the CP-violating $\theta$-term, 
\be
    \frac{\theta}{2\pi} \, \int \dd^4 x \, \tr \bigl( F \wedge F \bigr)\,,
\ee
where $F$ refers to the Yang-Mills field strength. This $\theta$-term contributes to the neutron electric dipole momentum. The strong CP-problem arises from the extremely tight experimental bound on the smallness of the dipole momentum, which acquires a very small $\theta$\,. The Peccei-Quinn approach to the strong CP problem promotes $\theta$ to a dynamical variable, \emph{i.e.}~the QCD axion. In this narrower context, the three-form naturally arises in the Peccei-Quinn theory: a non-zero $\theta$ implies that $\tr (F \wedge F)$ has a nonzero vacuum expectation value, while this topological term itself is the four-form field strength of a three-form gauge potential, identified with the Chern-Simons form. At high energies, however, new physics is expected to un-Higgs the three-form and a massless three-form coupled to the two-form dual of the axion has to be introduced. It is therefore tempting to ask whether our interpretation could also provide such an `un-Higgsed' theory for the QCD axion, which would require a different kinetic energy for the three-form potential rather than the one that we have in Eq.~\eqref{eq:S_massive}. However, at least in the M-theoretical embedding that we will present in Section~\ref{sec:mtu}, it appears that the axion in our construction is \emph{not} the QCD axion; instead, the axion as a brane-position mode should be thought of as an axion-like particle in a broader sense. 

\newpage

%%%%%%%%%%%%%%%%%%%%%%%%%%%%%%%%%%%%%%%%%
\subsection{Dimensional Reduction and Anti-Brane}\label{sec:back-reaction}
%%%%%%%%%%%%%%%%%%%%%%%%%%%%%%%%%%%%%%%%%

\begin{figure}
    \centering
    \hspace{5mm}
    \includegraphics[scale=0.65]{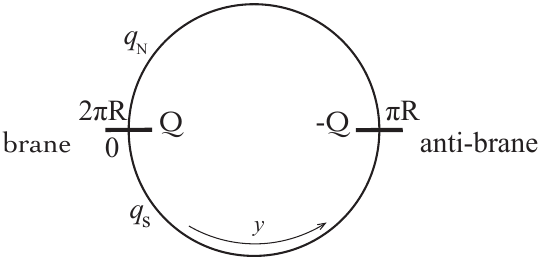} 
    \caption{An illustration of the antipodal two-brane configuration. Our universe as a 3-brane sits at $y=0$ and its anti-brane at $y = \pi \, R$\,, with $y$ the compactified dimension that is orthogonal to the brane. These two branes carry the same charge $\CQ$ but with opposite signs. The background value of the four-form field strength $\bar{H}_{01234}$ is a discontinuous function in $y$, which is equal to $q_\text{S}$ in the southern semi-circle and $q_\text{N}$ in the northern semi-circle.}
    \label{fig:circle}
\end{figure}

Here we provide a more self-contained setting that takes into account the back-reaction of the brane charges on the bulk fluxes. As a result, we will show that the axion potential in the four-dimensional EFT becomes multi-valued, similar to models of axion monodromy. As we will see, the self-consistency of the boundary condition in the compact direction requires that we generalize the action \eqref{eq:5D_full} to include an anti 3-brane with opposite charge. For now, we ignore gravitational effects but will get back to the stability of our setting later. We also assume that they have the same tension $\CT$ and are placed at $y_1=0$ and $y_2 = R \, \pi$ in the compact dimension. The five-dimensional action~\eqref{eq:5D_full} is now generalized to be
\begin{align*}
    S = - \frac{1}{2\cdot 5!}  \int_{\CM_5} \de^5 X \, H^{}_\mathrm{ABCDE} \, H^\mathrm{ABCDE}
    - \sum_{i = 1,\,2} \left\{ \mathcal{T} \int_{\CM_4^{(i)}} \de^4 x \, \sqrt{- \det \gamma } + \mathcal{Q}_i \int_{\CM_4^{(i)}} C^{(4)} \right\},
\end{align*}
where $\mathcal{Q}_1=-\mathcal{Q}_2 \equiv \mathcal{Q}$\,.
To proceed with the dimensional reduction, we decompose the four-form into a background and perturbation,
\begin{align}
C^{(4)} = \bar{C}^{(4)}(y) + \delta C^{(4)}\bigl(x\bigr)\,,
\end{align}
where only the background field depends on $X^4 = y$\,, which is compactified on a circle, with $y \sim y + 2\pi R$\,. Without loss of generality, we assume that the brane is located at $y^{}_1 = 0$ and the anti-brane at $y^{}_2 = \pi\,R$\,. The equation of motion from varying the action with respect to the background field $\bar{C}^{}_{0123}$ is
\begin{align}\label{eq:bulk_eom}
    \partial^{}_y \bar{H}^{}_{01234} = \CQ^{}_1 \, \delta \bigl( y - y^{}_1 \bigr) + \CQ^{}_2 \, \delta \bigl( y - y^{}_2 \bigr)\,.
\end{align}
Integrating this equation over the circle yields $\mathcal{Q}_1 + \mathcal{Q}_2=0$\,, which is what we have imposed. This implies the necessity of introducing the anti-brane.
We integrate Eq.~\eqref{eq:bulk_eom} to obtain
\begin{align}
    \bar{H}^{}_{01234} (y) = 
    \begin{cases}
        q^{}_\text{N} \quad \mathrm{for} \quad \pi R < y < 2\pi R\,, \\[4pt]
        q^{}_\text{S} \quad\, \mathrm{for} \quad\,\, 0 < y < \pi R\,,
    \end{cases}
        \qquad%
    q^{}_\text{S} = q^{}_\text{N} + \CQ\,.
\end{align}
This solution breaks the isometry in $y$ as $\bar{C}_{0123}$ now develops a dependence on $y$. 
For the fluctuations, we define 
$\delta H_{\alpha\beta\gamma\delta} \equiv \delta H_{4\alpha\beta\gamma\delta}$ and
$\delta C_{\alpha\beta\gamma} \equiv \delta C_{4\alpha\beta\gamma}$\,, 
such that 
$\delta H_{\alpha\beta\gamma\delta} = - 4 \, \partial_{[\alpha}\delta C_{\beta \gamma\delta]}$\,.
We are now ready to perform a dimensional reduction in $y$ to derive the effective action in four dimensions. This procedure follows analogously as in the previous subsection. The difference is that there are now two axions from both the brane and anti-brane, which we refer to as $a^{}_1$ and $a^{}_2$\,, respectively. The action~\eqref{eq:dim_reduced_action} now generalizes to
\begin{align} \label{eq:ca}
\begin{split}
    S^{(2)} = &-\frac{1}{2} \int  \de^4 x \Bigl( \tfrac{1}{4!} \, \delta H_\mathrm{\mu\nu\rho\sigma}\, \delta H^\mathrm{\mu\nu\rho\sigma} + \p_\mu a^{}_1 \, \p^\mu a^{}_1 + \p_\mu a^{}_2 \, \p^\mu a^{}_2 \Bigr) \\[4pt]
    & + \frac{1}{4!} \int \dd^4 x \, \epsilon^{\mu\nu\rho\sigma} \, \delta H^{}_{\mu\nu\rho\sigma} \, m \, \bigl( a^{}_1 - a^{}_2 - a^{}_0 \bigr) + S^{(2)}_\text{rest}\,, \\[4pt]
\end{split}
\end{align}
where
\be
    a_0 = \sqrt{\CT} \, \pi R \, \frac{q^{}_\text{N} + q^{}_\text{S}}{2 \, \CQ}\,,
\ee
and
\begin{align}
\begin{split}
    S^{(2)}_\text{rest} = & \frac{1}{2} \int \dd^4 x \bigg\{ \pi R \, \bigl( q^2_\text{N} + q^2_\text{S} \bigr) - 2 \, \p_\mu \Bigl[ \tfrac{1}{3!} \, \epsilon^{\mu\nu\rho\sigma} \delta C^{}_{\nu\rho\sigma} \, m \, \bigl( a^{}_1 - a^{}_2 \bigr) \Bigr] \biggr\} \\[4pt]
    & - \frac{\CQ}{4! \sqrt{\CT} \, R} \, \int \dd^4 x \, \epsilon^{\mu\nu\rho\sigma} \bigl( a^{}_1 - a^{}_2 \bigr) \, \p^{}_y \bar{C}^{}_{\mu\nu\rho\sigma}\,, 
\end{split}
\end{align}
In $S^{(2)}_\text{rest}$\,, the first line does not make any contribution to the dynamics of the axions and we thus drop it in the following discussion. In the second line, we encounter a term linear in $a^{}_1 - a^{}_2$\,. The coefficient of this latter linear terms is proportional to $\p_y \bar{C}_{0123} = \bar{H}_{01234}$\,, which is an undetermined constant on the brane itself. Following the rest of the derivations in Section~\ref{sec:mgtfp} by dualizing $\delta C^{}_{\mu\nu\rho}$\,, we find the analog of Eq.~\eqref{eq:ma}, with
\be \label{eq:facq}
    S^{(2)} \bigl[a^{}_1\,, a^{}_2\bigr] = - \frac{1}{2} \int \de^4 x \, \Bigl[ \p_\mu a^{}_1 \, \p^\mu a^{}_1 + \p_\mu a^{}_2 \, \p^\mu a^{}_2 + m^2 \, \bigl( a^{}_1 - a^{}_2 - a^{}_0 - \text{c} \bigr)^2  \Bigr],
\ee
where `c' denotes the undetermined constant shift due to the presence of the $\p_y \bar{C}_{\mu\nu\rho\sigma}$ term in the last line of Eq.~\eqref{eq:ca}. We can redefine away the dependence on `c' by \emph{e.g.} shift $a^{}_1$ as $a^{}_1 \rightarrow a^{}_1 + \text{c}$\,. However, the dependence on $a_0$ has an important implication for a discrete gauge symmetry, for which we will give a heuristic argument momentarily in this subsection.

We now identify the mass eigenstates by performing a change of basis,
\be
    a = \frac{a^{}_1 - a^{}_2}{\sqrt{2}}\,,
        \qquad%
    \tilde{a} = \frac{a^{}_1 + a^{}_2}{\sqrt{2}}\,.
\ee
It follows that the action~\eqref{eq:facq} becomes
\be \label{eq:facqme}
    S^{(2)} \bigl[a\,, \tilde{a}\bigr] = - \frac{1}{2} \int \de^4 x \, \Bigl[ \p_\mu a \, \p^\mu a + 2 \, m^2 \, \bigl( a - \tfrac{1}{\sqrt{2}} \, a^{}_0 \bigr)^2 \Bigr] - \frac{1}{2} \int \dd^4 x \, \p_\mu \tilde{a} \, \p^\mu \tilde{a}\,. 
\ee 
We therefore identify the mode $a$ as the axion, which has the mass
\be
    \sqrt{2} \, m = \frac{\CQ}{\sqrt{\pi \, R \, \CT}}\,.
\ee
The other mode $\tilde{a}$, however, remains massless. Geometrically, the massive mode corresponds to a relative brane displacement. The massless mode, by contrast, describes a shift of the center of mass. By construction, this EFT is only valid in a regime where the relative brane displacement is smaller than the size of the extra dimension, {\it i.e.} $|a| \lesssim   R\,\sqrt{\CT}$, which for a sufficiently large extra dimension $R$, or small axion mass equivalently, is a weak bound.

\begin{figure}[t!]
    \centering
    \hspace{5mm}
    \includegraphics[scale=1]{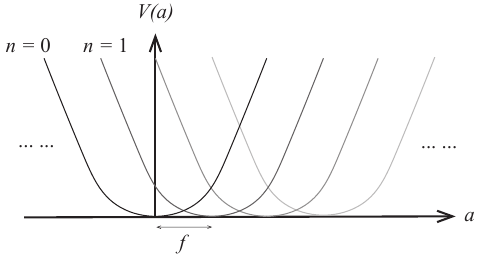} 
    \caption{The potential of the axion is multivalued, satisfying a discrete (residual) gauge symmetry with periodicity $f$.}
    \label{fig:quadpot}
\end{figure}

We now study the Dirac quantization condition associated with the background field strength $\bar{H}^{(5)}$\,. Quantum mechanically, if there exists a magnetic dual of the 3-brane, which is indeed the case in string theory, this brane charge has to be quantized. Imagine that we embed our five-dimensional EFT within ten-dimensional type IIB superstring theory, with the extra five directions compactified. As we will see later in Section~\ref{sec:mtu}, $H^{(5)}$ should be identified with the Ramond-Ramond four-form in IIB supergravity, and it has to satisfy the self-duality condition $H^{(5)} = \star H^{(5)}$, where `$\star$' denotes the Hodge dual in ten dimensions. Taking this string embedding seriously, without worrying about various subtleties regarding \emph{e.g.} supersymmetry breaking and the details of the compactification, we conclude that the total electric brane charge is
\be
    \mu \equiv \int \star \bar{H}^{(5)} = \pi R \, \CV \, \bigl( q^{}_\text{N} + q^{}_\text{S} \bigr) = \frac{2 \, a_0 \CV}{\sqrt{\CT}} \, \CQ \,,
\ee
with $\pi R \, \CV$ characterizing the volume of the compact hypersurface encircling the 3-brane in ten dimensions. 
The Dirac quantization condition then implies that $\mu = 2 \pi n / g$ is quantized, with $n \in \mathbb{Z}$ and $g$ the magnetic charge.   
This schematic argument shows that 
\be \label{eq:fn}
    a^{}_0 = \sqrt{2} \, f \, n\,,
        \qquad%
    f \equiv \frac{\pi \sqrt{\CT / 2}}{g \CV \CQ}\,,
        \qquad%
    n \in \mathbb{Z}\,,
\ee
which further implies that the axion theory~\eqref{eq:facq} admits a discrete gauge symmetry,\,\footnote{Strictly speaking, this is not a symmetry but instead a transformation between different superselection sectors.} 
\be
    a \rightarrow a + f \, k\,,
        \qquad%
    a^{}_0 \rightarrow a^{}_0 + \sqrt{2} \, f \, k\,,
        \qquad%
    k \in \mathbb{Z}\,. 
\ee
This discrete transformation makes the theory jump between different branches, labeled by $n$ in Eq.~\eqref{eq:fn}, in the potential depicted in Figure~\ref{fig:quadpot}. 
At the EFT level, this residual symmetry protects the axion potential from being spoiled by higher-order corrections and thus avoids any quality problem. This provides a physical picture that underlies the 
EFT construction 
in~\cite{Kaloper:2008fb, Kaloper:2011jz}.  Note that our construction is different from standard axion monodromy~\cite{Silverstein:2008sg, McAllister:2008hb}:  Instead of identifying the axion with a wrapped Kalb-Ramond or Ramond-Ramond potential, our axion arises from a brane-position mode. A detailed comparison with the standard treatment can be found in Appendix~\ref{sec:ccsam}.

Two massive branes attract each other gravitationally. A fully quantitative analysis would require solving the coupled system involving the five-dimensional gravitational field, the higher-form flux, and a possibly non-trivial dilaton profile, but here we limit ourselves to preliminary remarks. Define $d$ to be the distance in $y$ between the brane and anti-brane. If the extra dimension is stabilized with size $2 \pi R$, symmetry dictates that $V_\mathrm{gravity}'(d)$ vanishes in the antipodal configuration $d= \pi R$, where $d$ is the brane and anti-brane separation along $y$, because neither hemisphere is preferred (see also Fig.~\ref{fig:circle}). Moreover, for $d \ll R$ the compactness of the extra dimension becomes irrelevant and $V_\mathrm{gravity}(d)$ should scale linearly with $d$, analogous to the Coulomb potential between two charged plates.\,\footnote{See related discussions in~\cite{Burgess:2001fx}.} %These requirements alongside the invariance of $V_\mathrm{gravity}(d)$ under $d \to -d$ are then satisfied by
 A simple gravitational potential that satisfies these requirements alongside the invariance of $V_\mathrm{gravity}(d)$ under $d \to -d$ is given by
\begin{align} \label{eq:gp}
    V_\text{gravity}(d) =  C_1\,\frac{\CT^2}{2\pi R\,M_\mathrm{5}^3}  \, |d| \, (2\pi R - |d|),\quad d \in (-\pi R, \pi R]\,,
\end{align}
where $M_5^3$ is the five-dimensional Planck mass and $C_1$ an order unity constant.\,\footnote{The potential~\eqref{eq:gp} can be thought of as solving the Poisson equation $M_5^3\,V''_\text{gravity}(d) \propto \delta (d) - (2\pi R)^{-1}$, where the constant term $(2\pi R)^{-1}$ is required for consistency as can be seen by integrating both sides of the equation over the compact $y$-direction. Mathematically, it is possible to modify the constant to be a periodic function in the Poisson equation, but this would drastically change the background. Another possibility is that one of the branes should have a negative tension.}

%Here we give some preliminary remarks on the stability of this brane-anti-brane system.\,\footnote{See related discussions in~\cite{Burgess:2001fx}.} Schematically, the gravitational potential $V_\mathrm{gravity}$ in the fifth dimension should satisfy $M_5^3\,V''_\text{gravity}(d) \propto \delta (d) - (2\pi R)^{-1}$, where $M_5^3$ is the five-dimensional Planck mass, $d$ is the distance in $y$ between the brane and anti-brane, and the constant term $(2\pi R)^{-1}$ is required for consistency as can be seen by integrating both sides of the equation over the compact $y$-direction. Ultimately, this term should arise from the gravitational back-reaction of the branes. It follows that 
%
%\begin{align}
%V_\text{gravity}(d) =  C_1\,\frac{\CT^2}{2\pi R\,M_\mathrm{5}^3}  \, |d| \, (2\pi R - |d|),\quad d \in (-\pi R, \pi R]\,,
%\end{align}
%
%where $C_1$ an order unity constant. As a cross-check, $V_\mathrm{gravity}'$ vanishes in the antipodal configuration where $d= \pi R$ as expected due to the symmetry of the system under $d \to -d$. 
On the other hand, the axion mass gives rise to an effectively repulsive `force' described by the quadratic potential 
\begin{align}
V_\text{axion} (d) = C_2 \,  \frac{\CQ^2}{2\pi R} \, (d - d_0)^2\,,
\end{align}
where $C_2$ is another dimensionless constant. Expanding both potentials around the antipodal configuration at $d=\pi R$, we find that the (stable) axion potential dominates over the (tachyonic) gravitational one if $\CQ^2 \gtrsim  \CT^{\,2} M_5^{-3}$, which is independently required by the weak gravity conjecture for 3-brane gauge fields~\cite{Arkani-Hamed:2006emk,Ibanez:2015fcv, Palti:2020mwc,Kaloper:2023vrs}. At the same time, due to Eq.~\eqref{eq:axion_mass}, a sub-Planckian axion mass requires $\CQ^2 \ll 2 \pi R\, \CT\, M_\mathrm{pl}^2  $. Using $M_\mathrm{pl}^2 \sim 2 \pi R \, M_5^3$, we see that both conditions are compatible for $ \CT \ll M_\mathrm{pl}^4 $. In other words, within our simple EFT it appears possible to stabilize the brane position while keeping the axion light and satisfying the weak gravity conjecture.   Whether these competing forces could ultimately lead to a stable configuration within string theory requires a more detailed model building, which we leave for future studies.\footnote{In superstring theory, D-branes (of like-charges) are BPS (\emph{i.e.}~the brane charge saturates its mass) objects and exert no force on each other when static, as the NSNS (including gravity) and RR contributions exactly cancel each other~\cite{Polchinski:1995mt}. Applied to our case, the brane and anti-brane would therefore attract each other due to both gravity and gauge contributions to the potential. This might modify the stability condition derived here in a string theory embedding. In contrast, in the five-dimensional EFT setup, the higher-form gauge field is topological and its effect is captured by the axion due to Eq.~\eqref{eq:legendre1}.}

\vspace{3mm}

In summary, we have shown that a charged 3-brane in five dimensions 
provides a simple UV model of an axion with quadratic potential. In this picture, the pseudo-scalar character of the axion is inherited from the invariance of the higher-dimensional theory under parity transformations. Moreover, what appears as an approximately broken shift symmetry of the axion in the EFT is a natural consequence of the translational invariance of the brane in the fifth extra dimension. Flux quantization then renders the axion potential multi-valued. At the same time, this construction provides a simple gauge protection of its small mass and makes contact with a dual formulation in terms of a two-form field. While the five-dimensional model is self-contained as an EFT, the obvious question is of course if such a model can be embedded in string theory. In particular, we are interested whether the appearance of the three-form flux as well as the axion's dual formulation in terms of a two-form can be understood at a more fundamental level. We provide a first answer in the next section.

%%%%%%%%%%%%%%%%%%%%%%%%%%%%%%%%%%%%%%%%%
\section{An M-Theory Uplift} \label{sec:mtu}
%%%%%%%%%%%%%%%%%%%%%%%%%%%%%%%%%%%%%%%%%

We now embed the EFT construction in Section~\ref{sec:eftp} within the framework of string and M-theory, which provides further insights for the proposed axion model. We have discussed two dual descriptions of the axion. First, in Section~\ref{sec:afbb}, the axion is, by definition, a scalar satisfying the global shift symmetry. We will see that the (3+1)-dimensional brane-world considered therein receives an interpretation as the Dirichlet 3-brane (\emph{i.e.}~D3-brane) in string theory, with the axion being a brane-position mode that is an open string excitation on the D3-brane. Second, in Section~\ref{sec:catfd}, the axion is dualized to be a two-form gauge potential. It is less straightforward to interpret what this dual theory means in string theory. However, uplifted to M-theory in eleven dimensions, we will find that the dual two-form resides on the one-dimensional boundary of an `open' membrane, which is a two-dimensional extended object. This is reminiscent of how a one-form (Yang-Mills) field resides on the boundary of an open string, whose ends are particle-like. 

In Section~\ref{sec:mgtfp}, we introduced a three-form $C^{(3)}$, which played a central role in the mass generation~\cite{Dvali:2022fdv}. In the string theory uplift, $C^{(3)}$ arises as components of the Ramond-Ramond four-form in type IIB superstring theory. Although the IIB theory only admits even-degree Ramond-Ramond forms, the three-form arises due to the presence of a spatial compactification that is transverse to the D3-brane. This is because a T-duality transformation in the compact circle maps the IIB to IIA theory, where the latter contains odd forms. We will then further uplift this picture to M-theory, where $C^{(3)}$ becomes the gauge potential coupled to a `closed' membrane in M-theory. This is analogous to that a closed string couples to a Kalb-Ramond two-form, or that a particle couples to a one-form electromagnetic potential. 

We will also see that the M-theory 5-brane (\emph{i.e.}, the M5-brane), whose compactification over a two-torus gives rise to the 3-brane~\cite{Berman:1998va, Berman:1998sz}, plays a central role in the whole construction. The formulation of the M5-brane is rather involved. In view of its intriguing inspiration for axions, we find it timely to first give a pedagogical introduction to the M5-brane, before we connect to the EFT perspective later in this section. 

We will focus on the single-brane case in this section, where the theory is abelian. A non-abelian generalization naturally arises from considering a stack of coinciding branes, which does not change our discussion at the conceptual level. Moreover, even though all the string and branes in this section are supposedly supersymmetric, we will only consider their bosonic sectors, which are sufficient for our purpose. We will mostly focus on the matter sector, without considering the gravitational side of the story. This string theoretical embedding should be treated as an inspiration for the EFT construction of the axion model in Section~\ref{sec:eftp}, at least at the moment when it is still not clear whether (meta-stable) de Sitter space belongs to the string landscape or swampland~(see \emph{e.g.}~\cite{Danielsson:2018ztv, Cicoli:2018kdo, Obied:2018sgi, Agrawal:2018own} for modern debates on this subject). 

%%%%%%%%%%%%%%%%%%%%%%%%%%%%%%%%%%%%%%%%%
\subsection{Ingredients of M-Theory}
%%%%%%%%%%%%%%%%%%%%%%%%%%%%%%%%%%%%%%%%%

We start with a brief review of essential ingredients in M-theory~\cite{Hull:1994ys, Townsend:1995kk, Witten:1995ex}, which will facilitate our discussion on the M5-brane later. See \emph{e.g.}~\cite{Becker:2006dvp} for a more in-depth introduction. In the simplest setup, M-theory arises from ten-dimensional type IIA superstring theory in the strongly coupled regime, where the large string coupling $g^{}_\text{s}$ gives rise to an emergent eleventh dimension. This extra dimension is compactified over a circle of radius $R^{}_\text{s} \sim g^{}_\text{s} / \sqrt{T}$\,,
where $T$ is the string tension. In the limit where $g^{}_\text{s} \rightarrow \infty$\,, the circle decompactifies and we are led to eleven-dimensional M-theory. 

\vspace{3mm}

\noindent $\bullet$~\emph{M2-brane.} The fundamental role played by the one-dimensional string in string theory is now replaced with the two-dimensional membrane in M-theory, which is referred to as the M2-brane and propagates in the eleven-dimensional ambient spacetime~\cite{Bergshoeff:1987cm}.\,\footnote{Quantum mechanically, the supermembrane is unstable and tends to decay into small bits~\cite{deWit:1988xki}. One way to understand the dynamics of the supermembranes at the quantum level is via matrix theory~\cite{deWit:1988wri, Banks:1996vh}.} The bosonic sector of a single M2-brane with tension $T_\text{M2}$ is described by the action,  
\be \label{eq:mtba}
    S^{}_\text{M2} = - T_\text{M2} \int_{\Sigma_3} \dd^3 \sigma \, \sqrt{- \det \Bigl( \p^{}_{\mathbb{a}} X^{\mathbb{M}}_{\phantom{\dagger}} \, \p^{}_{\mathbb{b}} X^{\mathbb{N}}_{\phantom{\dagger}} \, \eta^{}_{\mathbb{M}\mathbb{N}} \Bigr)}\,,
        \qquad%
    \mathbb{a}\,,\,\mathbb{b} = 0\,, \, 1\,, \, 2\,,
\ee
where the embedding coordinates $X^\mathbb{M}$\,, $\mathbb{M} = 0\,, \, \cdots, \, 10$ map the three-dimensional worldvolume $\Sigma_3$ to the eleven-dimensional target space. We have introduced $\sigma^\mathbb{a}$ as the worldvolume coordinates on $\Sigma_3$\,. When there is a boundary $\p\Sigma_3$\,, we have an open M2-brane, which couples to an antisymmetric two-form gauge potential $\mathbb{A}^{}_{\mathbb{M}\mathbb{N}}$ via a Wess-Zumino term 
\be \label{eq:omgp}
    T_\text{M2} \int_{\p \Sigma_3} \mathbb{A}^{(2)}\,.
\ee
Here, the two-form 
$\mathbb{A}^{(2)}$
is pulled back from the target space to the boundary of the M2-brane. We continue to use the superscript `$(2)$' to indicate that $\mathbb{A}^{(2)}$ is a two-form as well as for other higher-form fields.

\vspace{3mm}

\noindent $\bullet$~\emph{Dimensional reduction to fundamental string.} In order to understand the physics of the M2-brane action, we first examine its dimensional reduction to the fundamental string in ten-dimensional superstring theory. Wrapping the M2-brane over a small circular target space direction $X^{10}$ gives rise to a string in ten dimensions. In practice, we set $X^{10} = \sigma^2$\,, under which the M2-brane action~\eqref{eq:mtba} becomes
\be \label{eq:stringa}
    S_\text{string} = - T \int_{\Sigma_2} \dd^2\sigma \, \sqrt{-\det \Bigl( \p^{}_a X^\text{M}_{\phantom{\dagger}} \, \p^{}_b X^\text{N}_{\phantom{\dagger}} \, \eta^{}_\text{MN} \Bigr)} + T \int_{\p\Sigma_2} A^{(1)}\,,
        \qquad%
    a = 0\,, \, 1\,,
\ee
with the worldsheet coordinates $\sigma^a$ and the embedding coordinates $X^\text{M}$, $M = 0 \, ,\, \cdots,\, 9$\,. This procedure is called a \emph{double-dimensional reduction} in the literature, as one dimensionally reduces a spatial direction on the M2-brane together with an ambient spatial direction. 

\vspace{3mm}

\noindent $\bullet$~\emph{D-branes.} In Eq.~\eqref{eq:stringa}, we have introduced a gauge potential $A^{(1)}$ that is pulled back from the target space to the boundary of the string, and it arises from open string excitations. Note that the ends of an open string have to reside on an extended object called D(irichlet)-brane~\cite{Horava:1989vt, Dai:1989ua, Horava:1989ga}, and the `$\mu$' index characterizes the directions that are longitudinal to the D-brane, along which the ends of the open string can move freely. In other words, the open string satisfies the Neumann boundary condition in these directions, and the ends of the open string are coupled to the gauge field $A^{(1)}$, whose components lies along the D-brane. In contrast, along the directions that are orthogonal to the D-brane, the open string satisfies the Dirichlet boundary condition and is coupled to the brane-position mode that perturb the shape of the D-brane. Such scalar modes arise from the spontaneous breaking of the translational symmetries in the directions transverse to the D-brane. In the string worldsheet action~\eqref{eq:stringa}, the dependence on such Nambu-Goldstone bosons are kept implicit, as they are part of the embedding coordinates in the unbroken phase. 

When the worldvolume of the D-brane is a ($p$+1)-dimensional manifold, which we denote as $\CM_{p+1}$\,, we have a D$p$-brane. The Dirichlet boundary condition is
$X^\text{M} |_{\CM_{p+1}} = f^\text{M} (x)$\,,
where we have introduced the worldvolume coordinates $x^\mu$\,, $\mu = 0\,, \, \cdots, \, p$\,. In static gauge, we align the brane with $X^\mu$ in the target space, such that
\be
    f^\mu = x^\mu\,, 
        \qquad%
    f^i = X^i_0 + \pi^i\,,
        \qquad%
    i = p + 1\,, \, \cdots, \, 9\,,
\ee
with $\pi^i$ the brane-position modes that fluctuate orthogonally to the D$p$-brane. Here, $X_0^i$ specify where the D$p$-brane is located in the transverse directions.

We now return to the one-form gauge potential $A^{(1)}$ in the string action~\eqref{eq:stringa}. We use the timelike coordinate $t$ to parametrize the particle-like boundary of the string, such that
$A^{(1)} = \dd t \, \p_t X^{\mu} A_\mu$\,.
The gauge potential $A_\mu$ lives on the D$p$-brane. Schematically, $A_\mu$ is related to the component of the two-form gauge potential $\mathbb{A}^{(2)}$ in the open membrane term~\eqref{eq:omgp} via 
$A_\mu = \mathbb{A}_{\mu\,, \, 10}$\,, with 
\be
    \mathbb{A}^{(2)} = \mathbb{A}_{\mu\nu} \, \dd X^\mu \wedge \dd X^\nu + 2 \, \mathbb{A}_{\mu\,, \, 10} \, \dd X^\mu \wedge \dd X^{10}\,.
\ee
Intuitively, the gauge coupling to $A^{(1)}$ is akin to the particle case, as the ends of an open string is effectively a point particle. 

In the following, we will mostly focus on the D3-brane, where the worldvolume is (3+1)-dimensional. The worldvolume action for a single D3-brane is\,\footnote{Note that for simplicity we assume that the dilaton is a constant, which we absorb into the brane tension.}
\be \label{eq:sdt}
    S_\text{D3} = - T_\text{D3} \int_{\CM_4} \dd^4 x \, \sqrt{-\det \Bigl( \p_\mu f^\text{M} \, \p_\nu f^\text{N} \, \eta^{}_\text{MN} + F_{\mu\nu} \Bigr)}\,,
\ee
where $\mu = 0\,, \, \cdots, \, 3$ and $F_{\mu\nu} = \p_\mu A_\nu - \p_\nu A_\mu$ is the $U(1)$ gauge field strength on the D3-brane. Moreover, $T_{\text{D}3}$ is the D3-brane tension. Here, $A_\mu$ originates from the open string modes that we have seen in Eq.~\eqref{eq:stringa}. These open strings on the D3-brane carry Chan-Paton charges at their ends, and they can move freely within the brane.    
Note that we have ignored the dependence on the dilaton (which is effectively set to a constant) for simplicity of the discussion. We will stick to this simplification through the rest of the paper. 

\vspace{3mm}

\noindent $\bullet$~\emph{M5-brane.} Now, we return to the discussion of the M2-brane~\eqref{eq:mtba}. Similar to the string case, the open membrane 
must also end on an extended object. There are not so many choices in M-theory: there exists simply one another extended object called M5-brane~\cite{Duff:1990xz, Gueven:1992hh}, which is a five-dimensional brane that arises from the magnetic dual of the M2-brane.~\footnote{Note that M2-brane can also end on another M2-brane, which we do not consider here.} Considering the scenario where M2-branes end on the M5-brane gives rise to the action for a single M5-brane in the Pasti-Sorokin-Tonin (PST) formulation~\cite{Pasti:1997gx, Pasti:1995tn, Pasti:1996vs},\footnote{A systematic way to motivate the M5-brane action~\eqref{eq:mfa0} is by dualizing the U(1) gauge potential on a single D4-brane in type IIA superstring theory~\cite{Aganagic:1997zk}. This procedure leads to a double dimensional reduction of the M5-brane action~\eqref{eq:mfa0} with a gauge-fixed PST scalar $b$. The six-dimensional Lorentz symmetry on the M5-brane worldvolume in this formulation is not manifest~\cite{Perry:1996mk, Schwarz:1997mc, Aganagic:1997zq}. In contrast, the PST formulation is manifestly Lorentz invariant due to the introduction of the extra scalar $b$\,.}
\be \label{eq:mfa0}
    S_\text{M5} = - T_\text{M5} \int_{\CM_6} \dd^6 x \left[ \sqrt{-\det\Bigl( \mathbb{G}_{\mathbb{m}\mathbb{n}} + i \, \mathbb{\Theta}_{\mathbb{m}\mathbb{n}} \Bigr)} + \frac{1}{4} \, \sqrt{-\mathbb{G}} \, \mathbb{\Theta}^{\mathbb{m}\mathbb{n}} \, \mathbb{F}_{\mathbb{m}\mathbb{n}\mathbb{k}} \, \mathbb{N}^\mathbb{k} \right], 
\ee
where $\mathbb{m} = 0\,, \, \cdots, \, 5$\,, $\mathbb{F}^{(3)} = \dd\mathbb{A}^{(2)}$\,, and
\be \label{eq:defmf}
    \mathbb{G}^{}_{\mathbb{m}\mathbb{n}} = \p_\mathbb{m} f^\mathbb{M} \, \p_\mathbb{n} f^\mathbb{N} \, \eta^{}_\mathbb{MN}\,,
        \qquad%
    \mathbb{N}^{}_\mathbb{m} = \frac{\p_\mathbb{m} b}{\sqrt{\p_\mathbb{n} b \, \mathbb{G}^{\mathbb{n}\mathbb{k}} \, \p_\mathbb{k} b}}\,,
        \qquad%
    \mathbb{\Theta}^{\mathbb{m}\mathbb{n}} = \frac{\epsilon^{\mathbb{m}\mathbb{n}\mathbb{k}\mathbb{l}\mathbb{p}\mathbb{q}} \, \mathbb{F}_{\mathbb{k}\mathbb{l}\mathbb{p}} \, \mathbb{N}_\mathbb{q}}{3! \, \sqrt{-\mathbb{G}}}\,.
\ee
Note that $\mathbb{G} = \det (\mathbb{G}_{\mathbb{m}\mathbb{n}})$\,, $\mathbb{G}^{\mathbb{m}\mathbb{n}}$ is the inverse of $\mathbb{G}_{\mathbb{m}\mathbb{n}}$\,, and the indices in the action are raised (or lowered) by the pullback metric $\mathbb{G}^{\mathbb{m}\mathbb{n}}$ (or $\mathbb{G}_{\mathbb{m}\mathbb{n}}$).
The embedding function $f^\mathbb{M}$ is given by
$f^{\mathbb{M}} = X^\mathbb{M} |_{\CM_6}$\,,
which specifies how the M5-brane is embedded within the eleven-dimensional target space. Again, we only focus on the bosonic part of the M5-brane action here and omit all the fermionic contents for simplicity of the presentation. 
The first term in the M5-brane action is in form similar to the D3-brane action~\eqref{eq:sdt}: in both cases, we have a determinant acting on the sum of the metric and gauge potential contribution. However, now that we have a three-form gauge field strength in the case where open M2-branes end on the M5-brane, one has to invoke the Hodge-dual quantity $\Theta^{\alpha\beta}$ in the M5-brane action, which replaces the two-form gauge field strength in the D3-brane action. The second term in Eq.~\eqref{eq:mfa0} represents a new feature of the M5-brane, which is required by supersymmetry in M-theory. 

One other important consequence of supersymmetry is that the two-form gauge potential $\mathbb{A}^{(2)}$ must satisfy a self-dual constraint on the M5-brane. This constraint is enforced by the auxiliary field $b$\,, called the PST scalar, that we have introduced in Eq.~\eqref{eq:defmf} via the definition of $\mathbb{N}_\mathbb{m}$\,. We now illustrate this self-dual constraint by focusing on the terms in the M5-brane action that are quadratic in the field strength $\mathbb{F}^{(3)}$\,. In the static gauge $\mathbb{G}_{\mathbb{m}\mathbb{n}} = \eta_{\mathbb{m}\mathbb{n}}$\,, these quadratic terms take the form, 
\be \label{eq:quada}
    - T_\text{M5} \int \dd^6 x \, \left\{ \frac{1}{24} \, \mathbb{F}^{\mathbb{m}\mathbb{n}\mathbb{l}} \, \mathbb{F}_{\mathbb{m}\mathbb{n}\mathbb{l}} - \frac{\p_\mathbb{k} b \, \p^\mathbb{l} b}{8 \, \p_\mathbb{p} b \, \p^\mathbb{p} b} \, \Bigl[ \mathbb{F}^{\mathbb{m}\mathbb{n}\mathbb{k}} - \bigl( \star \mathbb{F} \bigr)^{\mathbb{m}\mathbb{n}\mathbb{k}} \Bigr] \Bigl[ \mathbb{F}_{\mathbb{m}\mathbb{n}\mathbb{l}} - \bigl( \star \mathbb{F} \bigr)_{\mathbb{m}\mathbb{n}\mathbb{l}} \Bigr] \right\}.
\ee
Here, `$\star$' denotes the Hodge star on the six-dimensional worldvolume, with
\be
    \bigl( \star \mathbb{F} \bigr)^{\mathbb{m}\mathbb{n}\mathbb{k}} = \frac{1}{3!} \, \epsilon^{\mathbb{m}\mathbb{n}\mathbb{k}\mathbb{l}\mathbb{p}\mathbb{q}} \, \mathbb{F}^{}_{\mathbb{l}\mathbb{p}\mathbb{q}}\,. 
\ee
The equation of motion from varying the quadratic action~\eqref{eq:quada} with respect to the PST scalar imposes the self-dual constraint,
$\mathbb{F}^{(3)} = \bigl( \star \mathbb{F} \bigr)^{(3)}$.
The constrained two-potential $\mathbb{A}^{(2)}$ is said to be \emph{chiral}. 

The PST action for the M5-brane admits extra gauge symmetries in addition to the one-form gauge symmetry of the chiral gauge field $\mathbb{A}^{(2)}$, \emph{i.e.}~$\delta_\xi \mathbb{A}^{(2)} = \dd \xi^{(1)}$\,. These extra gauge symmetries are referred to as the PST symmetries, which act on the PST scalar as 
\be \label{eq:pstgauge}
    \delta^{}_\text{PST} b = \varphi\,.
\ee
The gauge symmetry~\eqref{eq:pstgauge} implies that $b$ is not dynamical and can be eliminated by fixing the gauge. In addition to the zero-form PST symmetry parametrized by $\varphi$\,, there is an additional one-form PST symmetry that acts on the chiral field $\mathbb{A}^{(2)}$ as
\be \label{eq:pstt}
    \delta^{}_\text{PST} \mathbb{A}^{(2)} = \frac{1}{2} \, \dd b \wedge \chi^{(1)} + g(\varphi)\,,
\ee
where $\chi^{(1)}$ is a one-form gauge parameter. The action of the zero-form PST symmetry on $\mathbb{A}^{(2)}$ is rather involved, which we hide within the unspecified function $g(\varphi)$ and refer the readers to~\cite{Pasti:1997gx} for details. 

\subsection{Toroidal Compactification of the M5-Brane} \label{sec:tcmfb}

We now derive a D3-brane from compactifying the M5-brane, which has a (5+1)-dimensional worldvolume, over a two-torus. See~\cite{Berman:1998va, Berman:1998sz} for the original work on this compactification.\,\footnote{See also Section~4.1 in~\cite{Ebert:2023hba} for a review, whose convention we follow closely here. Note that \cite{Ebert:2023hba} considers a generalization that manifests the SL($2,\mathbb{Z}$) symmetry. In contrast, we focus on the simpler case where the PST scalar $b$ is fixed and the SL($2,\mathbb{Z}$) symmetry becomes hidden.} We will see that, at the lowest order, this M-theoretical construction coincides with the proposal of the EFT of an axion in Section~\ref{sec:eftp}. Denote the coordinates on the two-torus as $x^\text{m}$\,, $\text{m} = 4\,, \, 5$\,, which by definition satisfy the periodic boundary conditions $x^4 \sim x^4 + 1$ and $x^5 \sim x^5 + 1$\,.
For simplicity, let the line element on the torus be
$\dd s^2 = \Gamma \, \dd x^\text{m} \, \dd x^\text{m}$\,, such that the area of the torus is $\Gamma$\,.
In order to perform the dimensional reduction, we take $X^9 = x^4$ and $X^{10} = x^5$\,. 
Note that this is a double-dimensional reduction with respect to both coordinates, \emph{i.e.}~we compactify two brane spatial directions together with the same ambient spatial directions.
To describe the wrapping of the M5-brane around the torus, we take the following (simplified) ansatz for the metric $\mathbb{G}_{\mathbb{m}\mathbb{n}}$ in the action~\eqref{eq:mfa0}, which is a pullback from the target spacetime to the M5-brane worldvolume:
\be \label{eq:gabkkao}
    \mathbb{G}^{}_{\mathbb{m}\mathbb{n}} =
    \begin{pmatrix}
        \mathbb{G}^{}_{\mu\nu} &\,\, 0 \\[4pt]
        0 &\,\, \Gamma \, \delta^{}_\text{mn}
    \end{pmatrix},
        \qquad%
    \mathbb{G}^{}_{\mu\nu} = \p^{}_\mu f^{\text{M}'} \, \p^{}_\nu f^{\text{N}'} \, \eta^{}_{\text{M}'\text{N}'}\,.
\ee
Here, $\text{M}' = 0\,, \cdots\,, 8$\,, as the target space becomes nine-dimensional after the compactifying over a two-torus. Note that, for simplicity, we have ignored the Kaluza-Klein vectors here, which would not change the discussion at the conceptual level. 

Now, we fix the PST gauges. We first fix the zero-form PST gauge~\eqref{eq:pstgauge} such that $b = \sigma^5$\,, with $\sigma^5$ the fifth spatial direction on the M5-brane. Then, we fix the one-form PST gauge~\eqref{eq:pstt} such that $\mathbb{A}^{}_\text{mn} = 0$\,. We further define
\be \label{eq:fj}
    F^{}_{\mu\nu} = \mathbb{F}^{}_{\mu\nu 4}\,,
        \qquad%
    J^\mu = \frac{\epsilon^{\mu\nu\rho\sigma} \, \mathbb{F}^{}_{\nu\rho\sigma}}{3! \, \sqrt{- \mathbb{G}}}\,. 
\ee
Double-dimensionally reducing the M5-brane action on the two-torus gives rise to the following four-dimensional worldvolume action:
\begin{align} \label{eq:dra}
    S = - T_\text{M5} \, \Gamma \int_{\CM_4} \dd^4 x \, \sqrt{-\det \Bigl[ \mathbb{G}^{}_{\mu\nu} + i \, \Gamma^{-1/2} \, \bigl( \star F^{(2)} \bigr)^{}_{\mu\nu} -  J^{}_\mu \, J^{}_\nu \Bigr]}\,. 
\end{align}
Note that a total derivative term has been dropped. 
Here, the indices are lowered by $G_{\mu\nu}$\,. 
This theory provides a non-linear generalization of the action~\eqref{eq:S_2form} of the two-form gauge field that is dual to the axion. In order to make this connection manifest, we go to the static gauge $G{}^{}_{\mu\nu} = \eta^{}_{\mu\nu}$ and set $F^{(2)} = 0$ for now, while $J_\mu$ is taken to be nonzero. Expand the action with respect to a small $J_\mu$\,, at the quadratic order we find
\be \label{eq:fdquad}
    - \frac{1}{12} \, T_\text{M5} \, \Gamma \int_{\CM_4} \dd^4 x \, \mathbb{F}^{\mu\nu\rho} \, \mathbb{F}^{}_{\mu\nu\rho}\,,
\ee
where $\mathbb{F}^{}_{\!\mu\nu\rho} = \p^{}_\mu \mathbb{A}^{}_{\nu\rho} + \p^{}_\nu \mathbb{A}^{}_{\rho\mu} + \p^{}_\rho \mathbb{A}^{}_{\mu\nu}$\,. As expected, the four-dimensional action~\eqref{eq:fdquad} is part of the first term in the six-dimensional action~\eqref{eq:quada}. In Section~\ref{sec:catfd}, we then performed an electromagnetic duality of the two-form gauge potential $\mathbb{A}_{\mu\nu}$ to find the axion field in the dual frame. We now perform the same electromagnetic duality of $\mathbb{A}_{\mu\nu}$\,, but in the full bosonic M5-brane action~\eqref{eq:dra}. 

In order to perform the electromagnetic duality transformation, we treat $\mathbb{F}_{\mu\nu\rho}$ in $J^\mu$ from Eq.~\eqref{eq:fj} as an independent field (instead of an exact form), but modify the action~\eqref{eq:dra} to be the following `parent' action:
\be \label{eq:sparent}
    S_\text{parent} = S[\mathbb{F}_{\mu\nu\rho}] + S_\text{g.f.}[\mathbb{F}_{\mu\nu\rho}\,,\,\mathbb{A}_{\mu\nu}\,,\,\pi_\mu]\,,
\ee
with the generating functional,
\be \label{eq:gfworr}
    S_\text{g.f.} = \frac{T_\text{M5}}{3!} \, \int_{\CM_4} \dd^4 x \, \epsilon^{\mu\nu\rho\sigma} \, \pi_\mu \, \Bigl( \mathbb{F} - d\mathbb{A} \Bigr)_{\nu\rho\sigma}\,. 
\ee
Integrating out the auxiliary field $\pi_\mu$ in the associated path integral imposes the condition $\mathbb{F}_{\mu\nu\rho} = \bigl( \dd\mathbb{A} \bigr){}_{\mu\nu\rho}$\,, which leads us back to the original action~\eqref{eq:dra}. Instead, we now integrate out $\mathbb{A}_{\mu\nu}$ in the path integral associated with the parent action~\eqref{eq:sparent}, which imposes the exactness condition $\dd \pi^{(1)} = 0$\,, where $\pi^{(1)} = \pi_\mu \, \dd x^\mu$\,. This condition can be solved locally by
\be \label{eq:pmmp2}
    \pi_\mu = \p_\mu \pi\,,
\ee
where $\pi$ is a scalar field. Plugging Eq.~\eqref{eq:pmmp2} back into the original parent action~\eqref{eq:sparent}, followed by integrating out $\mathbb{F}_{\mu\nu\rho}$\,, we find the following dual action:
\be \label{eq:dualafd}
    S_\text{dual} = - T_\text{M5} \, \Gamma \int_{\CM_4} \dd^4 x \sqrt{- \det \Bigl( \mathbb{G}^{}_{\mu\nu} + \Gamma^{-2}_{\phantom{\dagger}} \, \p_\mu \pi \, \p_\nu \pi + F^{}_{\mu\nu} \Bigr)}\,.  
\ee
In order to obtain this action, we performed a rescaling $F^{(2)} \rightarrow \sqrt{\Gamma} \, F^{(2)}$. The ambient spacetime arises from compactifying eleven-dimensional M-theory over a two-torus. Na\"{i}vely, in the limit where the torus shrinks to zero with $\Gamma \rightarrow 0$\,, one expects to get a nine-dimensional target space. However, this is incorrect intuition as it misses a stringy effect. From the definition~\eqref{eq:gabkkao} of the metric $G^{}_{\mu\nu}$\,, we see that it is a pullback from a nine dimensional subspace to the four-dimensional worldvolume. In addition, the scalar field $\pi(x)$ in Eq.~\eqref{eq:dualafd} perturbs the shape of the resulting 3-brane in an extra `emergent' $X^9$ direction that is orthogonal to the brane. We are thus led to a 3-brane in ten-dimensional ambient spacetime, which is made more manifest by rewriting the dual action as
\be
    S_\text{D3} = - T_\text{D3} \int_{\CM_4} \dd^4 x \sqrt{- \det \Bigl( \p^{}_\mu f^{\text{M}} \, \p^{}_\nu f^{\text{N}} \, \eta^{}_{\text{M}\text{N}} + F^{}_{\mu\nu} \Bigr)}\,,
\ee
with $\text{M} = 0\,, \, \cdots, \, 9$\,, $f^9 = y^{}_0 + R \, \pi (x)$\,, $R = \Gamma^{-1}$, and the tension $T_\text{D3}=T_\text{M5} \, \Gamma $\,. Note that we have set the fundamental length scale to one here. In this form, the dual action receives an interpretation as a D3-brane in type IIB superstring theory (see discussions around Eq.~\eqref{eq:sdt}), with $R$ the radius of the compact direction that is transverse to the D3-brane. In the decompactification limit where $R \rightarrow \infty$ we recover type IIB superstring theory in ten dimensions~\cite{Schwarz:1995dk, Aspinwall:1995fw}.~\footnote{A heuristic argument for why compactifying eleven-dimensional M-theory over a two-torus that shrinks to zero leads to ten-dimensional type IIB superstring theory is given below: Take one of the cycles of the two-torus to be the M-theory circle, compactifying on which leads to type IIA superstring theory. Then, we T-dualize the remaining cycle of the two-torus to get type IIB superstring theory. It then follows that shrinking the original cycle in the IIA theory decompactifies the dual circle, which leads to IIB theory in ten dimensions.}
This D3-brane is localized at $y_0$ in the $X^9$ direction, which is compactified over a circle of an effective radius $R$\,. We observe that the effective radius $R$ of the $X^9$ circle is identified with the inverse area of the two-torus, over which the M5-brane is compactified. At the quadratic order in $\pi (x)$\,, we recover the free scalar theory~\eqref{eq:fa}, where $\pi (x)$ is interpreted as a massless axion field.  

\subsection{Three-From Potential from M-Theory}

In Section~\ref{sec:mgtfp} we have seen that, in order for the axion to acquire a small mass, it is crucial that the axion is coupled to a three-form $C^{(3)}$~\cite{Dvali:2005an}. We now explain how such a three-form gauge field arises in the framework of M-theory. 

\vspace{3mm}

\noindent $\bullet$~\emph{Strings and $B$-field.} Let us first return to the action~\eqref{eq:stringa} describing the fundamental string, which does not yet include all possible terms, even only when the bosonic sector is concerned. It is also possible to turn on a Wess-Zumino term in the string sigma model,
\be \label{eq:swz}
    S^\text{WZ}_\text{string} = \frac{T}{2} \int_{\Sigma_2} \dd^2 \sigma \, \epsilon^{ab} \, \p^{}_a X^\mu \, \p^{}_b X^\nu \, B^{}_{\mu\nu}\,, 
\ee
where $\epsilon^{ab}$ is the Levi-Civita symbol satisfying $\epsilon^{01} = 1$ and $B^{}_{\mu\nu}$ the anti-symmetric Kalb-Ramond two-form field. In a constant $B$-field, it is manifest that this Wess-Zumino term is purely topological: it is only non-vanishing when the string winds around a compact spatial direction in the target space of the sigma model. To be specific, we compactify $X^1$ over a circle of radius $R$\,, such that
$X^1 \bigl( \sigma + 2\pi \bigr) = X^1 \bigl( \sigma \bigr) + w \, R \, \sigma$\,,
with $w \in \mathbb{Z}$\,,
where we have normalized the length of the string to be $2\pi$\,. This means that, when we traverse along the string once, the $X^1$-circle in the target space is traversed over $w$ times, and that the string has winding number $w$\,. In static gauge with $X^0 = \tau$ and in the presence of a nonzero $B_{01}$\,, we find
$S_\text{string}^\text{WZ} = T \, w \, R \, B^{}_{01} \, \mathcal{V}_\text{string}$\,,
where $\mathcal{V}_\text{string}$ is the spacetime volume of the string. This is reminiscent of the $\theta$-term in QCD, which captures the instanton number that counts how many times a three-sphere wraps around another three-sphere. 

\vspace{3mm}

\noindent $\bullet$~\emph{D-brane and Ramond-Ramond potentials.} A similar Wess-Zumino term can also be added to the D-branes. For example, the D3-brane action~\eqref{eq:sdt} admits the Wess-Zumino term,
\be
    S_\text{D3}^\text{WZ} = \frac{T_\text{D3}}{4!} \int \dd^4 \sigma \, \epsilon^{\mu\nu\rho\sigma} \, \p^{}_\mu f^\text{M} \, \p^{}_\nu f^\text{N} \, \p^{}_\rho f^\text{K} \, \p^{}_\sigma f^\text{L} \, C^{}_\text{MNKL}\,,
\ee
where $C^{}_\text{MNKL}$ is a Ramond-Ramond four-form potential. In terms of the pulled back differential form defined on the D3-brane worldvolume,
$C^{(4)} \equiv \frac{1}{4!} \,\dd f^\text{M} \! \wedge \dd f^\text{N} \! \wedge \dd f^\text{K} \! \wedge \dd f^\text{L} \, C^{}_\text{MNKL}$\,,
we write $S_\text{D3}^\text{WZ} = T_\text{D3} \int C^{(4)}$\,.
In perturbative string theory, the D-brane dynamics is encoded by the fundamental string excitations that perturb the brane. Therefore, we expect that the $B$-field, which arises from certain closed string excitations, must show up somewhere in the D-brane action. Such dependence on the $B$-field can be inferred by examining the open string contribution in the string worldsheet action~\eqref{eq:stringa}, which we rewrite as
$\int_{\p\Sigma_2} A^{(1)} = \int_{\Sigma_2} F^{(2)}$,
with
$F^{(2)} = \dd A^{(1)}$.
We observe that the $B$-field in Eq.~\eqref{eq:swz} can be combined with the $U(1)$ gauge field strength $F^{(2)}$ to form the action term 
\be
    \int_{\Sigma_2} \CF^{(2)}\,,
        \qquad%
    \CF^{(2)} = B^{(2)} + F^{(2)}\,. 
\ee
Therefore, introducing the dependence on the $B$-field amounts to a simple replacement of $F_{\mu\nu}$ with $\CF_{\mu\nu}$ in the D3-brane action~\eqref{eq:sdt}. Moreover, just like how the D3-brane couples to the lower-degree differential form $\CF^{(2)}$ originated from the fundamental string (other than the four-form $C^{(4)}$), it also couples to Ramond-Ramond potentials associated with the lower-dimensional branes, which can be smeared within the D3-brane~\cite{Douglas:1995bn}. Recall that the D3-brane lives in type IIB superstring theory, where there are only even-degree Ramond-Ramond potentials. The ones of degree lower than four include $C^{(2)}$ coupled to the D1-string and $C^{(0)}$ coupled to the D(-1)-brane, which is an instanton. 
The only sensible way to turn on such couplings in the D3-brane action is by forming the following degree-four differential forms: 
$C^{(2)} \wedge \CF^{(2)}$ and $C^{(0)} \, \CF^{(2)} \wedge \CF^{(2)}$.
The bosonic sector of the D3-brane worldvolume action that generalizes Eq.~\eqref{eq:sdt} is
\begin{align} \label{eq:sdtg}
\begin{split}
    S_\text{D3} & = - T_\text{D3} \int_{\CM_4} \dd^4 x \, \sqrt{-\det \Bigl( G^{}_{\mu\nu} + \CF_{\mu\nu} \Bigr)} \\[4pt]
    & \quad + T_\text{D3} \int_{\CM_4} \Bigl( C^{(4)} + C^{(2)} \wedge \CF^{(2)} + \tfrac{1}{2} \, C^{(0)} \, \CF^{(2)} \wedge \CF^{(2)} \Bigr)\,,
\end{split}
\end{align}
with $G^{}_{\mu\nu} \equiv \p^{}_\mu f^\text{M} \, \p^{}_\nu f^{}_\text{M}$\,. Setting $f^\mu = x^\mu$ and at the quadratic order in small $F^{(2)}$\,, we find
\be
    S^\text{quad.}_\text{D3} = - \frac{T_\text{D3}}{4} \int_{\CM_4} \dd^4 x \, \Bigl( F^{}_{\mu\nu} \, F^{\mu\nu} - \tfrac{1}{2} \, \epsilon^{\mu\nu\rho\sigma} \, C^{(0)} \, F^{}_{\mu\nu} \, F^{}_{\rho\sigma} \Bigr)\,.
\ee
Note that there is \emph{no} linear term in $F_{\mu\nu}$ because it is an anti-symmetric tensor.
Here, the second term is related to the $\theta$-coupling associated with the QCD instanton,\,\footnote{In the literature, it is standard to refer to $C^{(0)}$ as an axion~\cite{Becker:2006dvp}.} which is distinct from the axion-like particle as a brane-position mode that we are proposing in this paper -- the latter is an open string excitation.

\vspace{3mm}

\noindent $\bullet$~\emph{Higher-form gauge potentials in M-theory.} Analogous to the Wess-Zumino term~\eqref{eq:swz} in the string sigma model, where the string is coupled to the Kalb-Ramond two-form $B^{(2)}$, a membrane in eleven-dimensional M-theory is naturally coupled to a three-form $\mathbb{C}^{(3)}$. The complete bosonic sector of the M2-brane action that generalizes Eq.~\eqref{eq:mtba} is~\cite{Bergshoeff:1987cm}
\be
    S^{}_\text{M2} = - T_\text{M2} \int_{\Sigma_3} \dd^3 \sigma \, \sqrt{- \det \Bigl( \p^{}_{\mathbb{a}} X^{\mathbb{M}}_{\phantom{\dagger}} \, \p^{}_{\mathbb{b}} X^{\mathbb{N}}_{\phantom{\dagger}} \, \eta^{}_{\mathbb{M}\mathbb{N}} \Bigr)} + T_\text{M2} \int_{\Sigma_3} \mathbb{C}^{(3)}\,,
\ee
Similarly, analogous to the D3-brane action~\eqref{eq:sdtg}, we generalize the M5-brane action~\eqref{eq:mfa0} to include couplings to the three-form $\mathbb{C}^{(3)}$ as~\cite{Pasti:1997gx}
\begin{align} \label{eq:mfa}
    S^{\phantom{k}}_\text{M5} & = S^\text{kin.}_\text{M5} \bigl[ \mathscr{F}^{(3)} \bigr] + T_\text{M5} \int_{\CM_6} \Bigl( \mathbb{C}^{(6)} + \mathbb{C}^{(3)} \wedge \mathbb{F}^{(3)} \Bigr)\,,   
\end{align}
where
\be \label{eq:skinmf}
    S^\text{kin.}_\text{M5} \bigl[ \mathscr{F}^{(3)} \bigr] = - T_\text{M5} \int_{\CM_6} \dd^6 x \left[ \sqrt{-\det\Bigl( \mathbb{G}_{\mathbb{m}\mathbb{n}} + i \, \mathbb{\Theta}_{\mathbb{m}\mathbb{n}} \Bigr)} + \sqrt{-\mathbb{G}} \, \mathbb{\Theta}^{\mathbb{m}\mathbb{n}} \, \mathscr{F}_{\mathbb{m}\mathbb{n}\mathbb{k}} \, \mathbb{N}^\mathbb{k} \right],
\ee
and
\be
    \Theta^{\mathbb{m}\mathbb{n}} = \frac{\epsilon^{\mathbb{m}\mathbb{n}\mathbb{l}\mathbb{p}\mathbb{q}} \, \mathscr{F}^{}_{\mathbb{k}\mathbb{l}\mathbb{p}} \, \mathbb{N}^{}_{\mathbb{q}}}{3! \, \sqrt{-\mathbb{G}}}\,,
        \qquad%
    \mathscr{F}^{(3)} = \mathbb{C}^{(3)} + \mathbb{F}^{(3)}\,.
\ee
Here, $\mathbb{C}^{(6)}$ is the six-form potential coupled to the M5-brane, just like how the Ramond-Ramond four-form $C^{(4)}$ is coupled to the D3-brane in Eq.~\eqref{eq:sdtg}. 

\vspace{3mm}

\noindent $\bullet$~\emph{M5-brane on a two-torus revisited.} In Section~\ref{sec:tcmfb} we have derived the compactification of the kinetic part~\eqref{eq:skinmf} over a two-torus, in the case where $\mathbb{C}^{(3)}$ is set to zero. Now, we discuss the case where the three-form gauge field $\mathbb{C}^{(3)}$ is included~\cite{Berman:1998va, Berman:1998sz}, whose pullback on the M5 gives rise to the following quantities after dimensionally reducing on a torus:
\be
    \mathbb{C}^{}_{\mu45} = C^{}_{\mu}\,,
        \qquad%
    \begin{pmatrix}
        \mathbb{C}^{}_{\mu\nu 4} \\[4pt]
        \mathbb{C}^{}_{\mu\nu 5}
    \end{pmatrix}
        = 
        \begin{pmatrix}
            B^{}_{\mu\nu} \\[4pt]
            C^{}_{\mu\nu}
        \end{pmatrix},
            \qquad%
    \mathbb{C}^{}_{\mu\nu\rho} = C^{}_{\mu\nu\rho}\,.
\ee
Here, in the resulting string theory, $B^{(2)}$ is identified as the Kalb-Ramond field and $C^{(q)}$ the Ramond-Ramond fields.  
In order to focus on the effect of $C^{(3)}$, which will play an important role in the mass-generation of the axion, we further set $C^{(1)} = B^{(2)} = C^{(2)} = \mathbb{C}^{(6)} = 0$ for simplicity.  We then find that the (double) dimensionally reduced action~\eqref{eq:dra} generalizes to
\be \label{eq:drat}
    S^{}_\text{d.r.} \bigl[ \mathscr{F}^{(3)} \bigr] = - T_\text{M5} \, \Gamma \int_{\CM_4} \dd^4 x \, \sqrt{-\det \! \left[ \mathbb{G}^{}_{\mu\nu} + i \, \frac{\bigl( \star F^{(2)} \bigr)^{}_{\!\mu\nu}}{\Gamma^{1/2}} -  \bigl( \star \mathscr{F}^{(3)} \bigr)_{\!\mu} \, \bigl( \star \mathscr{F}^{(3)} \bigr)_{\!\nu} \right]}\,,
\ee
Here, $\mathscr{F}^{(3)}$ is understood to have the components $\mathscr{F}^{}_{\mu\nu\rho}$ on the dimensionally reduced four-dimensional worldvolume. Note that there is \emph{no} additional Wess-Zumino term, simply because we have chosen to set most higher-forms to zero in order to focus on $C^{(3)}$ to start with. Effectively, the only difference between Eqs.~\eqref{eq:dra} and \eqref{eq:drat} is that $J^{}_\mu$ is now replaced with
\be
    \bigl( \star \mathscr{F}^{(3)} \bigr)^{}_\mu = \frac{\mathbb{G}^{}_{\mu\nu} \, \epsilon^{\nu\rho\sigma\lambda} \, \mathscr{F}^{}_{\rho\sigma\lambda}}{3! \, \sqrt{- \mathbb{G}}}\,,
        \qquad%
    \mathscr{F}^{}_{\mu\nu\rho} = C^{}_{\mu\nu\rho} + \mathbb{F}_{\mu\nu\rho}\,.
\ee
Next, we dualize $\mathbb{F}_{\mu\nu\rho}$ in the action.~\eqref{eq:drat}, which means constructing the `parent' action by adding in the same generating functional~\eqref{eq:gfworr}, such that
\be
    S^{}_\text{parent} = S^{}_\text{d.r.}  \bigl[ \mathscr{F}^{(3)} \bigr]+ \frac{T_\text{M5}}{3!} \, \int_{\CM_4} \dd^4 x \, \epsilon^{\mu\nu\rho\sigma} \, \pi^{\phantom{\dagger}}_\mu \, \Bigl( \mathbb{F} - \dd\mathbb{A} \Bigr){}^{\phantom{\dagger}}_{\nu\rho\sigma}\,,
\ee
where $\mathbb{F}^{}_{\mu\nu\rho}$ is now regarded as an independent field. Integrating out $\mathbb{A}^{}_{\mu\nu}$ imposes that $\pi_\mu = \p_\mu \pi$ as we have seen earlier. However, instead of directly integrating out $\mathbb{F}^{}_{\mu\nu\rho}$\,, it is convenient to first rewrite the parent action as
\be
    S^{}_\text{parent} = S^{}_\text{d.r.} \bigl[ \mathscr{F}^{(3)} \bigr] + \frac{T_\text{M5}}{3!} \, \int_{\CM_4} \dd^4 x \, \epsilon^{\mu\nu\rho\sigma} \, \p^{}_\mu \pi \, \mathscr{F}^{}_{\nu\rho\sigma} - T_\text{M5} \int_{\CM_4} \dd\pi \wedge C^{(3)}\,.
\ee
Now, we can equivalently integrate $\mathscr{F}^{}_{\mu\nu\rho}$ out instead. This procedure is in form identical to what we have see in Section~\ref{sec:tcmfb}. We therefore recover the dual D3-brane action~Eq.~\eqref{eq:dualafd}, but now also including the dependence on $C^{(3)}$\,, 
\be \label{eq:ddtbct}
    S_\text{D3} = - T^{}_{\text{D}3} \int_{\CM_4} \dd^4 x \sqrt{- \det \Bigl( \mathbb{G}^{}_{\mu\nu} + \Gamma^{-2}_{\phantom{\dagger}} \, \p^{}_\mu \pi \, \p^{}_\nu \pi + F^{}_{\mu\nu} \Bigr)} - T^{}_{\text{D}3} \, \int_{\CM_4} \Gamma^{-1} \, \dd\pi \wedge C^{(3)}\,.
\ee
%
%with $T_{\text{D}3} = T_{\text{M}5}$\,. 

\vspace{3mm}

\noindent $\bullet$~\emph{General D3-brane action.} Recovering the dependence on the higher-order gauge forms that we have omitted, the same double-dimensional reduction procedure leads to the full action on the four-dimensional worldvolume,
\begin{align} \label{eq:dtba}
\begin{split}
    S^{}_\text{D3} & = - T_{\text{D}3} \int_{\CM_4} \dd^4 x \, \sqrt{- \det \Bigl[ \mathbb{G}^{}_{\mu\nu} + \bigl( C^{}_\mu + R \, \p^{}_\mu \pi \bigr) \, \bigl( C^{}_\nu + R \, \p^{}_\nu \pi \bigr) + \CF^{}_{\mu\nu} \Bigr]} \\[4pt]
    & \quad - T_{\text{D}3} \int_{\CM_4} \Bigl[ \bigl( \mathbb{C}^{(4)} + R \, \dd\pi \wedge C^{(3)} \bigr) + C^{(2)} \wedge \CF^{(2)} + \tfrac{1}{2} \, C^{(0)} \, \CF^{(2)} \wedge \CF^{(2)} \Bigr]\,,
\end{split}
\end{align}
where $\CF^{(2)} = B^{(2)} + F^{(2)}$ and
$\mathbb{C}^{(4)} \equiv \frac{1}{4!} \, \dd f^{\text{M}'} \!\wedge \dd f^{\text{N}'} \!\wedge \dd f^{\text{K}'} \!\wedge \dd f^{\text{L}'} \, \mathbb{C}^{}_{\text{M}'\text{N}'\text{K}'\text{L}'}$,
$\text{M}' = 0\,, \, \cdots, \, 8$
comes from dimensionally reducing $\mathbb{C}^{(6)}$ in Eq.~\eqref{eq:mfa}. Moreover, $R = \Gamma^{-1}$ is the radius of the compact circle. This dimensionally reduced action becomes identical to Eq.~\eqref{eq:sdtg} in ten dimensions, upon identifying the following standard Kaluza-Klein Ans\"{a}tze:
\begin{subequations}
\begin{align}
    G^{}_\text{MN} \, \dd X^\text{M} \, \dd X^\text{N} &= G_{\text{M}'\text{N}'} \, \dd X^{\text{M}'} \, \dd X^{\text{N}'} + \Bigl( C_{\text{M}'} \, \dd X^{\text{M}'} + \dd X^9 \Bigr) \, \Bigl( C_{\text{N}'} \, \dd X^{\text{N}'} + \dd X^9 \Bigr)\,, \\[4pt]
    C^{(4)} &= \mathbb{C}^{(4)} + \dd X^9 \wedge C^{(3)}\,,
        \qquad%
    X^9 = y_0 + R \, \pi(x)\,,
\end{align}
\end{subequations}
with $y_0$ the location of the D3-brane in the $X^9$ circle and $C_{\text{M}'}$ the Kaluza-Klein vector. 

%%%%%%%%%%%%%%%%%%%%%%%%%%%%%%%%%%%%%%%%%
\subsection{Axion from M5-Brane}
%%%%%%%%%%%%%%%%%%%%%%%%%%%%%%%%%%%%%%%%%

We have already shown how various results obtained in this section are connected to the EFT perspective in Section~\ref{sec:eftp}. In this last subsection, we collect all the relevant threads and give a coherent UV perspective of what we have discussed in Section~\ref{sec:eftp} at the EFT level. 

Our starting point is the M5-brane action~\eqref{eq:mfa}, which we compactify over a two-torus to get the four-dimensional worldvolume action~\eqref{eq:drat}. At the quadratic order in 
$\mathscr{F}^{}_{\mu\nu\rho} = C^{}_{\mu\nu\rho} + \mathbb{F}^{}_{\mu\nu\rho}$\,, 
and up to a normalization factor, Eq.~\eqref{eq:drat} reproduces the first term in Eq.~\eqref{eq:fdquad0}, with $C^{}_{\mu\nu\rho}$ originating from components of the three-form $\mathbb{C}_{\mathbb{m}\mathbb{n}\mathbb{l}}$ coupled to the closed membrane in eleven dimensions before the compactification, and $\mathbb{F}^{}_{\mu\nu\rho}$ from the field strength $\mathbb{F}^{}_{\mathbb{m}\mathbb{n}\mathbb{l}}$ of the two-form gauge field $\mathbb{A}^{}_{\mathbb{m}\mathbb{n}}$ coupled to an open membrane ending on the M5-brane. However, we still miss the kinetic term $H^{}_{\mu\nu\rho\sigma} \, H^{\mu\nu\rho\sigma}$ in Eq.~\eqref{eq:fdquad0}, with $H^{(4)} = \dd C^{(3)}$ the field strength associated with the three-form potential. This kinetic term originates from the supergravity sector. The bosonic sector of eleven-dimensional supergravity is described by~\cite{Cremmer:1978km}
\be
    S^{}_{11} = \frac{1}{2 \, \kappa^2_{11}} \int \dd^{11} x \, \sqrt{-\mathbb{G}} \left[ R (\mathbb{G}) - \frac{1}{2 \cdot 4!} \, \mathbb{H}^{}_{\mathbb{MNLK}} \, \mathbb{H}^{\mathbb{MNLK}} \right] - \frac{1}{12 \, \kappa^2_{11}} \int \mathbb{C}^{(3)} \wedge \mathbb{H}^{(4)} \wedge \mathbb{H}^{(4)}\,,
\ee
where $\mathbb{H}^{(4)} = \dd \mathbb{C}^{(3)}$ and $\kappa^{}_{11}$ is the gravitational coupling in eleven dimensions. After compactifying over the two-torus, a projection of the $\mathbb{H}^{}_{\mathbb{MNLK}} \, \mathbb{H}^{\mathbb{MNLK}}$ term on the M5-brane gives rise to the desired kinetic term in Eq.~\eqref{eq:fdquad}. 

In the above, we explained the M-theory uplift of the two-form gauge field. In the frame where the two-form is dualized back to the axion as a scalar field, the dimensionally reduced M5-brane is dualized to the four-dimensional D3-brane, which is described by the action~\eqref{eq:ddtbct}. Expand up to the quadratic order of the axion $\pi$\,, we recover the second and third term in the EFT action~\eqref{eq:S_massive}. From the D3-brane perspective, the axion $\pi$ is the brane-position mode that perturbs the shape of the brane. Moreover, the three-form term arises from a dimensional reduction of the D3-brane charge term $\int C^{(4)}$\,. The reason is that there is an extra spatial compactification that is T-dual to one of the cycles of the toroidal compactification in M-theory. However, again, we miss the kinetic term $H^{}_{\mu\nu\rho\sigma} \, H^{\mu\nu\rho\sigma}$\,. As expected, this term comes from the closed string sector described at low energies by IIB supergravity. Focusing on the Ramond-Ramond potential $C^{(4)}$ and setting all the other gauge potentials to zero, the bosonic sector of IIB supergravity is described~\cite{Schwarz:1983qr, Howe:1983sra}:
\be \label{eq:tbsg}
    S^{}_\text{IIB} = \frac{1}{2 \, \kappa^2_{10}} \int \dd^{10} x \, \sqrt{-G} \, \left[ R (G) - \frac{1}{4 \cdot 5!} \, H^{}_\text{ABCDE} \, H^\text{ABCDE} \right].
\ee
Here, $H^{(5)} = \dd C^{(4)}$ and $\kappa^{}_{10}$ is the gravitational coupling in ten dimensions and the dilaton has been set to constant for simplicity. Moreover, $G_\text{MN}$ is the ten-dimensional metric, $G = \det (G_\text{MN})$\,, and $R(G)$ is the associated Ricci scalar. The action~\eqref{eq:tbsg} has to be supplemented with a self-duality constraint $H^{(5)} = \star H^{(5)}$\,, which cannot be easily incorporated using an action principle. For this reason, the original work~\cite{Schwarz:1983qr, Howe:1983sra} only used the field equations and supersymmetry transformations. The kinetic term for $C^{(4)}$ may be viewed as a ten-dimensional uplift of the first term in the five-dimensional action~\eqref{eq:5D_full}.  Projecting this term to the D3-brane and reading off the relevant components as in Section~\ref{sec:mgtfp}, we recover the kinetic term for the three-form potential $C^{(3)}$ in Eq.~\eqref{eq:S_massive}.

There is one important subtlety here. Note that, in Section~\ref{sec:mgtfp}, the mechanism generating a mass for the axion does not take into account any self-duality constraint for the five-form field strength. This means that supersymmetry has to be broken at the energy scale where the mass generation occurs. Supersymmetry breaking in string theory is of course required for any phenomenological applications, but it is also known to be difficult. Together with the choice of vacuum state in string theory landscape, these are the standard major challenges in string phenomenology. For these reasons, we are still far from a realistic UV-completion, at least at the current stage. However, at the very least, the M-theory uplift here provides further inspirations and intriguing insights for the EFT proposal of the axion in Section~\ref{sec:eftp}.

%%%%%%%%%%%%%%%%%%%%%%%%%%%%%%%%%%%%%%%%%
\section{Conclusions} \label{sec:concl}
%%%%%%%%%%%%%%%%%%%%%%%%%%%%%%%%%%%%%%%%%

In this paper we proposed an axion universe that was inspired by M-theory. In this proposal the two-form dual to the axion originates from the two-form gauge potential coupled to open membranes ending on an M-theory 5-brane. Compactifying over a two-torus, we are led to a universe on a D3-brane, where the axion becomes the brane-position mode that perturbs the shape of the D3-brane. The axion mass is generated via a Higgs mechanism for a three-form gauge potential, which is ultimately coupled to the closed membrane in M-theory. Guided by these inputs from string and M-theory, we were led to the EFT construction as detailed in Section~\ref{sec:eftp}, where the axion is the position mode of a (3+1)-dimensional brane in the ambient (4+1)-dimensional spacetime, and the global U(1) shift symmetry is a residue of the diffeomorphism in the fifth direction orthogonal to the brane. 

Our proposal provides a geometric interpretation via an M-theoretical embedding of the field-theory constructions in~\cite{Dvali:2005an,Dvali:2022fdv, Kaloper:2008fb, Kaloper:2011jz}. More broadly, our work points towards a new program of mapping out open string axions as brane-position modes of various branes. For example, it would be interesting to understand implications of the non-geometric effects associated with exotic branes for cosmology. While initial steps have been taken in this paper, many intriguing open questions still remain for future explorations. 

On the theoretical side, it would be interesting to further examine any imprints of the proposed embedding in M-theory on the EFT of the axion in Section~\ref{sec:eftp}. 
We have discussed the quantization condition associated with the Ramond-Ramond four-form potential coupled to the D3-brane in Section~\ref{sec:back-reaction}. 
The flux compactification typically requires a warped geometry, which introduces a mass for the scalar field controlling the size of the extra dimensions and hence stabilize the moduli fields. Such a warped geometry may lead to interesting twists of our EFT construction for the axion. Moreover, we have also 
mentioned the subtlety regarding the self-duality constraint on the five-form field strength in type IIB superstring theory. It would be important to investigate whether there is any imprint of this constraint at low energies, where supersymmetry is explicitly broken. 

On the phenomenological side, apart from the stability of the compactification that is closely related to the elements that we have commented on above, it would also be interesting to understand the implications of the brane-anti-brane configuration in more detail. In particular, in terms of our EFT model, this includes solving the full flux-gravity system taking into account the brane backreaction on the geometry. Moreover, in Eq.~\eqref{eq:axion_mass} we provide a perturbative expression for the axion mass. The fact that we have an explicit UV model for the axion will allow us to study potential non-perturbative contributions to the axion mass.\footnote{For example, as pointed out in~\cite{Dvali:2004tma} in the context of charged 2-branes, a bilinear term in the three-form potential might be generated through quantum corrections. Such a term could lead to (non-perturbative) corrections to the axion mass if generalized to our setting.} In Section~\ref{sec:back-reaction}, we have shown that in the case of a compact extra dimension, it is necessary to introduce a second brane with a negative charge due to the self-consistency of the flux associated with the three-form potential. 
Whether one could establish a (meta-)stable configuration that survives through the age of the Universe remains to be explored. 

\acknowledgments

We would like to thank Aleksandr Chatrchyan, Georgi Dvali, Nemanja Kaloper, Johannes Lahnsteiner, Niels A. Obers, Tom\'{a}s Ort\'{i}n, Antonio Padilla, and Martin S.~Sloth for useful discussions. 
The work of F.N. is supported by VR Starting Grant 2022-03160 of the Swedish Research Council.
The work of Z.Y. is supported in part by Olle Engkvists
Stiftelse Project Grant 234-0342, VR Project Grant 2021-04013, the European Union’s
Horizon 2020 research and innovation programme under the Marie Sklodowska-Curie Grant Agreement No.~31003710, and the Villum Young Investigator Programme under project No.~71589. The Center of Gravity is a Center of Excellence funded by the Danish National Research Foundation under grant No.~184.

\appendix

\section{Closed versus Open String Axion Monodromy} \label{sec:ccsam}

We emphasize that the M-theory inspired axion that we consider here is distinct from the common scenario where the axion is identified with a dual of the Kalb-Ramond two-form (or, more generically, Ramond-Ramond potentials)~\cite{Silverstein:2008sg, McAllister:2008hb}, where the latter is a closed string mode. The two-form that we consider here, which is dual to the axion, is instead coupled to the open membrane when uplifted to eleven dimensions. Compactified down to ten-dimensional string theory, part of this uplifted two-form in M-theory gives rise to the one-form gauge potential coupled to the fundamental open string, while the remaining degree of freedom is further dualized to be the axion as a brane-position mode on the four-dimensional worldvolume. Such brane-position modes are associated with the open instead of closed string.

The literature on string theory inspired axions is vast, which contain both closed and open string examples. Nevertheless, to our knowledge, the open string axion proposal considered in this paper still points towards a rather new (and simple) arena. Below we review some relevant theories of axion monodromy for comparison and to clarify the differences. For a pedagogical introduction, see \emph{e.g.}~\cite{Baumann:2014nda}.

\vspace{3mm}

\noindent $\bullet$~\emph{Linear potential from D5- and NS5-brane.} We consider the bosonic part of the D5-brane action, with \emph{no} Ramond-Ramond potential,\,\footnote{Note that the D5-brane also arises from compactifying an M5-brane, but now the compactification is over a two-torus with one cycle longitudinal to the brane and the other transverse to the brane. (This is distinct from the case considered in the bulk of the current paper, where both cycles are taken to be along the M5-brane.) Shrinking this two-torus leads to a D5-brane in ten-dimensional type IIB superstring theory.} 
\be \label{eq:dfb}
    S_\text{D5} = - T^{}_\text{D5} \int \dd^6 x \, \sqrt{-\det \bigl( G^{}_\text{MN} + B^{}_\text{MN} \bigr)}\,,
\ee
where we focus on the dynamics created by the Kalb-Ramond two-form $B^{(2)}$ by setting all the other field fluctuations to zero. Require that the D5-brane fill the four-dimensional spacetime $\CM_4$ of our Universe and wrap a two-cycle $\Sigma_2$ that is transverse to our Universe, such that
$b = \int_{\Sigma_2} B^{(2)}$ 
is identified with the axion. We also require that $B^{(2)}$ vanish when projected to $\CM_4$\,. Schematically, we assume that 
\be
    G^{}_\text{MN} = \eta^{}_\text{MN}\,,
        \qquad%
    B^{(2)} = \ell^{-2} \, b \bigl(x^0, \cdots, x^3\bigr) \, \dd x^4 \wedge \dd x^5,
        \qquad%
    \int_{\Sigma_2} \dd x^4 \wedge \dd x^5 \sim \ell^2\,.
\ee
In the large field regime where $b \gg \ell^2$\,, \emph{i.e.}~the characteristic size of the compact manifold $\Sigma_2$ is very small, the D5-brane action~\eqref{eq:dfb} becomes
$S_\text{D5} \sim - \int_{\CM_4} \! \dd^4 x \, b$\,,
which gives rise to a potential linear in the axion $b$\,. On the other hand, the supergravity contains the kinetic term associated with $B^{(2)}$, which leads to the following canonically normalized action describing an axion with a linear potential:
\be
    S^{}_\text{D5} \sim - \frac{1}{2} \int_{\CM_4} \!\! \dd^4 x \, \bigl( \p_\mu b \, \p^\mu b + \lambda \, b \bigr).
\ee
A similar construction also applies to the NS5-brane that is S-dual to the D5-brane, except that in this case the Kalb-Ramond field is traded with the Ramond-Ramond two-form, which now underlies the axion. This NS5 scenario is typically preferred due to the $\eta$-problem associated with the Kalb-Ramond axion. See~\cite{McAllister:2008hb} for further details. 

\vspace{3mm}

\noindent $\bullet$~\emph{Quadratic potential from D7-brane.} It was also proposed in~\cite{Palti:2014kza} that a quadratic potential can be obtained by considering a D7-brane compactified over a four-dimensional internal manifold $\Sigma_4$\,. The NSNS sector of the D7-brane action is
\be
    S^{}_\text{D7} = - T^{}_\text{D7} \int \dd^8 x \, \sqrt{-\det\bigl( \eta^{}_\text{MN} + B^{}_\text{MN} \bigr)}\,.
\ee
As a proof of concept, we now take the na\"{i}ve assumption that
\be
    B^{(2)} = \ell^{-2} \, b \bigl(x^0, \cdots, x^3\bigr) \, \bigl( \dd x^4 \wedge \dd x^5 + \dd x^6 \wedge \dd x^7 \bigr)\,,
        \qquad%
    \int_{\Sigma_4} \!\! \dd x^4 \wedge \cdots \wedge \dd x^7 = \ell^4\,.
\ee
Again, taking the large field limit, we find
$S^{}_\text{D7} \sim -\int_{\CM_4} \dd^4 x \, b^2$. 
Combined with the kinetic term from the supergravity sector, we are now led to an axion with a quadratic potential that, for example, underlies chaotic inflation. 

\vspace{3mm}

\noindent $\bullet$~\emph{Three-form.} Another proposal for realizing a quadratic potential is by coupling the axion to a four-form field strength, which generates an axion mass via a Higgs mechanism~\cite{Dvali:2005an, Kaloper:2011jz}. This consideration led to the EFT in form the same as the one that we have also described in Section~\ref{sec:mgtfp}. However, our construction in Section~\ref{sec:mgtfp} is distinct from what we have reviewed so far in this section. As we have emphasized, the kinetic term of our axion as a brane-position mode arises from the metric part of the brane action. This is an example of open string axions. This is in contrast to the closed string axions (\emph{e.g.}~as Kalb-Ramond or Ramond-Ramond modes) that are more commonly considered in the literature. Indeed, as we have demonstrated in this paper, the two-form dual of the axion can be identified with an open membrane mode, which is unrelated to either Kalb-Ramond or Ramond-Ramond modes.

\vspace{3mm}

\noindent $\bullet$~\emph{Wilson line.} In spirit, our proposal of open string axion is reminiscent of an open string axion monodromy modes considered in~\cite{Marchesano:2014mla}, even though the detailed construction is still very different. We will compare our construction with~\cite{Marchesano:2014mla} in the next bullet point. In order to understand the connection to~\cite{Marchesano:2014mla}, here
we consider the T-duality of our axion as a position mode of the D3-brane, which is interesting on its own. 

In the T-dual frame, the position mode is mapped to a Wilson line in the dual D4-brane. For pedagogical reason, we briefly review the derivation of this T-dual relation in the simplest setup~\cite{Polchinski:1998rr}. Consider a Wilson line on a D4-brane wrapping around a compact circle of radius $R$ in the $y$ direction,
\be
    W = \exp \bigl[ i \, q \, \theta (x) \bigr]\,,
        \qquad%
    \theta (x) = \oint \dd y \, A (x^\mu, \, y)\,.
\ee
For simplicity, we take $\theta (x)$ to be constant. 
For a charged particle, the presence of the gauge potential induces a shift in the canonical momentum $p^{}_y$ in $y$, with
$p_y^{} = n + q \, \theta / (2\pi R)$\,, $n \in \mathbb{Z}$\,.
Consider a neutral open string ending on two coinciding D4-branes carrying unit charges with the opposite signs and associated with the Wilson lines $\theta_1$ and $\theta_2$\,. It follows that
$p^{}_y = n + ( \theta^{}_1 - \theta^{}_2 ) / (2\pi R)$\,.
The zero mode of the associated embedding coordinate is then
$y \sim p^{}_y \, \tau$,
with $\tau$ the string worldsheet time. Under T-duality along the $y$ direction, a Neumann boundary condition is mapped to a Dirichlet boundary condition, which maps the D4-brane extending in $y$ to a D3-brane localized in $y$. The duality map is 
$\p^{}_\tau y = \p^{}_\sigma \tilde{y}$\,, 
with $\tilde{y}$ the T-dual direction compactified over a dual circle of radius $\tilde{R} \sim 1 / R$ and $\sigma \in [0\,, \, 2\pi)$ the spatial coordinate on the string worldsheet. Focusing on the zero mode, we have
\be \label{eq:dyt}
    \tilde{y} \sim \! \lr \tilde{w} + \frac{\theta_1 - \theta_2}{2\pi} \rr \! \tilde{R} \, \sigma\,, 
        \qquad%
    \tilde{w} = n \in \mathbb{Z}\,.
\ee
Here, $\tilde{w}$ is interpreted as the integer winding of the open string wrapping around the compact dimension. The second term in Eq.~\eqref{eq:dyt} is the `fractional' winding number, arising from the open string stretching between two D3-branes at $\theta_1 \, \tilde{R}$ and $\theta_2 \, \tilde{R}$\,, respectively. This is how the Wilson line on a D4-brane is T-dualized to the position of a D3-brane localized in the compact $y$ direction. In general, $\theta(x)$ can depend on the D3-brane coordinates. In the T-dual frame, this scalar corresponds to the brane-position mode of the D3-brane. This construction provides a T-dual description of the open string axion that we formulated in this paper. 

\vspace{3mm}

\noindent $\bullet$~\emph{Twisted torus.}
In contrast, an explicit example considered in~\cite{Marchesano:2014mla} starts with a D7-brane wrapped on a twisted three-torus with a massive Wilson line along the $S^1$ fiber. Performing a T-duality transformation along the fiber direction in the twisted torus gives rise to a D6-brane with the brane-position mode $a$ that acts as an axion. Moreover, in this T-dual frame, the twisted torus is mapped to a standard three-torus with a three-form flux $k \, \dd x^5 \wedge \dd x^6 \wedge \dd x^7$. Here, $x^5$, $x^6$, and $x^7$ are coordinates on the internal three-torus, with $x^7$ the fiber direction. This flux is  associated with a non-trivial Kalb-Ramond field $B^{(2)} = k \, x^7 \, \dd x^5 \wedge \dd x^6$, which breaks the translational invariance in the $S^1$ fiber direction in $x^7$. Let $x^7 = x^{}_0 + a$\,, with $x^{}_0$ the location of the localized D6-brane. We then consider the D6-brane action,
\be
    S^{}_\text{D6} = - T^{}_\text{D6} \int \dd^7 x \, \sqrt{-\det\bigl( G^{}_\text{MN} + B^{}_\text{MN} \bigr)}\,,
\ee
with the following schematic dimensional reduction ansatz for the pullback metric on the D6-brane,
\be
    G^{}_\text{MN} = 
    \begin{pmatrix}
        \eta^{}_{\mu\nu} + \p^{}_\mu a \, \p^\mu a &\,\, 0 \\[4pt]
        0 &\,\, \delta^{}_{ij}
    \end{pmatrix}\!,
        \quad%
    \mu = 0\,, \cdots\!, 3\,, 
        \quad
    i = 4\,, 5\,, 6\,,
\ee
where we have ignored all the modes that are irrelevant to our discussion here. 
Note that we dimensionally reduce the directions $x^4,\cdots,x^9$. 
Keeping up to the second order in $a$, we find
\be
    S^{}_\text{D6} \sim - \int_{\CM_4} \!\!\dd^4 x \, \Bigl[ \p_\mu a \, \p^\mu a + k^2 \, \bigl( a + x^{}_0 \bigr)^2 \Bigr]\,,
\ee
with the axion mass $k$\,. Even though we both realize the axion as an open string mode, the mass generating mechanism in~\cite{Marchesano:2014mla} involves a non-trivial Kalb-Ramond field and clearly differs from our framework in this paper. 

\vspace{3mm}

\noindent $\bullet$~\emph{Torsion homology.} In~\cite{Marchesano:2014mla}, there is also another interesting model that realizes the EFT considered in~\cite{Dvali:2005an, Kaloper:2011jz}, which is relevant to our discussion here. This construction requires a compactification that has a nontrivial torsion homology, which generalizes the idea of a twisted torus. Consider a higher-form gauge potential $A^{(n+1)}$ propagating in four-dimensional spacetime times an $n$-dimensional internal space. 
The Kaluza-Klein reduction of the gauge potential $A^{(n+1)}$ over the internal space is given by
\be
    A^{(n+1)} = B^{(2)} \wedge \sigma^{(n-1)} + C^{(3)} \wedge \lambda^{(n-2)}\,,
\ee
where $B^{(2)}$ is identified as the dual of the axion in four-dimensions. 
The structure of torsion homology implies that there is a torsion $(n-1)$-form $\sigma^{(n-1)}$ that is related to a $(n-2)$-form $\lambda^{(n-2)}$ via 
\be
    \dd \lambda^{(n-2)} = k \, \sigma^{(n-1)}.
\ee
The field strength of $A^{(n+1)}$ then gives
\be
    F^{(n+2)} \equiv \dd A^{(d+1)} = \bigl( \dd B^{(2)} - k \, C^{(3)} \bigr) \wedge \sigma^{(n-1)} + \dd C^{(3)} \wedge \lambda^{(n-2)}\,. 
\ee
Upon dimensional reduction to four-dimensions, we then find that
\be
    \int \dd^{n+4} x \, \bigl| F^{(n+2)} \bigr|^2 \sim \int \dd^4 x \, \Bigl( \bigl| \dd C^{(3)} \bigr|^2 + \mu \, \bigl| \dd B^{(2)} \! - k \, C^{(3)} \bigr|^2 \Bigr)\,. 
\ee
This is essentially identical to Eq.~\eqref{eq:fdquad0}. Intriguingly, this proposal in a sense provides a closed string realization for the EFT in~\cite{Dvali:2005an, Kaloper:2011jz}. 

\newpage

\bibliographystyle{JHEP}
\bibliography{atfmt}

\end{document}